\documentclass[conference]{IEEEtran}
\IEEEoverridecommandlockouts
\usepackage{cite}
\usepackage{amsmath,amssymb,amsfonts}
\usepackage{algorithmic}
\usepackage{graphicx}
\usepackage{hyperref}
\usepackage{textcomp}
\usepackage{xcolor}
\definecolor{Grey}{gray}{0.9}
\usepackage{colortbl}
\usepackage{booktabs}

\graphicspath{{./Figures/}}

\makeatletter
\newcommand{\linebreakand}{%
  \end{@IEEEauthorhalign}
  \hfill\mbox{}\par
  \mbox{}\hfill\begin{@IEEEauthorhalign}
}
\makeatother

\newlength{\figgap}
\def\BibTeX{{\rm B\kern-.05em{\sc i\kern-.025em b}\kern-.08em
    T\kern-.1667em\lower.7ex\hbox{E}\kern-.125emX}}

\begin{document}

\newif\ifanonymous
%\anonymousfalse

\ifanonymous
\title{Where do we Jam? - The Impact of Real and Virtual XR Spaces on Musical Creativity, Co-Presence, Affect, and Flow\\
\thanks{Funding information omitted for anonymous review.}
}
\else
\title{Where do we Jam? - The Impact of Real and Virtual XR Spaces on Musical Creativity, Co-Presence, Affect, and Flow\\
\thanks{This work was partially funded by the DFG through the Leibniz award of Elisabeth André (AN 559/10-1), and by the MUSMET project funded by the EIC Pathfinder Open scheme of the European Commission (grant agreement n. 101184379).}
}
\fi

\ifanonymous
\author{
\IEEEauthorblockN{Anonymous Author 1}
\IEEEauthorblockA{\textit{Anonymous Institution} \\
Anonymous City, Anonymous Country \\
anonymous@email.com}
\and
\IEEEauthorblockN{Anonymous Author 2}
\IEEEauthorblockA{\textit{Anonymous Institution} \\
Anonymous City, Anonymous Country \\
anonymous@email.com}
\linebreakand
\IEEEauthorblockN{Anonymous Author 3}
\IEEEauthorblockA{\textit{Anonymous Institution} \\
Anonymous City, Anonymous Country \\
anonymous@email.com}
\and
\IEEEauthorblockN{Anonymous Author 4}
\IEEEauthorblockA{\textit{Anonymous Institution} \\
Anonymous City, Anonymous Country \\
anonymous@email.com}
\and
\IEEEauthorblockN{Anonymous Author 5}
\IEEEauthorblockA{\textit{Anonymous Institution} \\
Anonymous City, Anonymous Country \\
anonymous@email.com}
}
\else
\author{
\IEEEauthorblockN{Ruben Schlagowski}
\IEEEauthorblockA{\textit{HCAI-Lab} \\
\textit{University of Augsburg}\\
Augsburg, Germany \\
ruben.schlagowski@uni-a.de}
\and
\IEEEauthorblockN{Jonas Bordewick}
\IEEEauthorblockA{\textit{HCAI-Lab} \\
\textit{University of Augsburg}\\
Augsburg, Germany \\
jonas.bordewick@uni-a.de}
\linebreakand
\IEEEauthorblockN{Luca Turchet}
\IEEEauthorblockA{\textit{Creative, Intelligent \& Multisensory} \\ 
\textit{Interactions Laboratory, University of Trento}\\
Trento, Italy\\
luca.turchet@unitn.it}
\and
\IEEEauthorblockN{Hannes Ritschel}
\IEEEauthorblockA{\textit{HCAI-Lab} \\
\textit{University of Augsburg}\\
Augsburg, Germany \\
hannes.ritschel@uni-a.de}
\and
\IEEEauthorblockN{Elisabeth André}
\IEEEauthorblockA{\textit{HCAI-Lab} \\
\textit{University of Augsburg}\\
Augsburg, Germany \\
elisabeth.andre@uni-a.de}
}
\fi
\maketitle

\begin{figure*}[ht!]
	\centering
	\includegraphics[width=.99\textwidth]{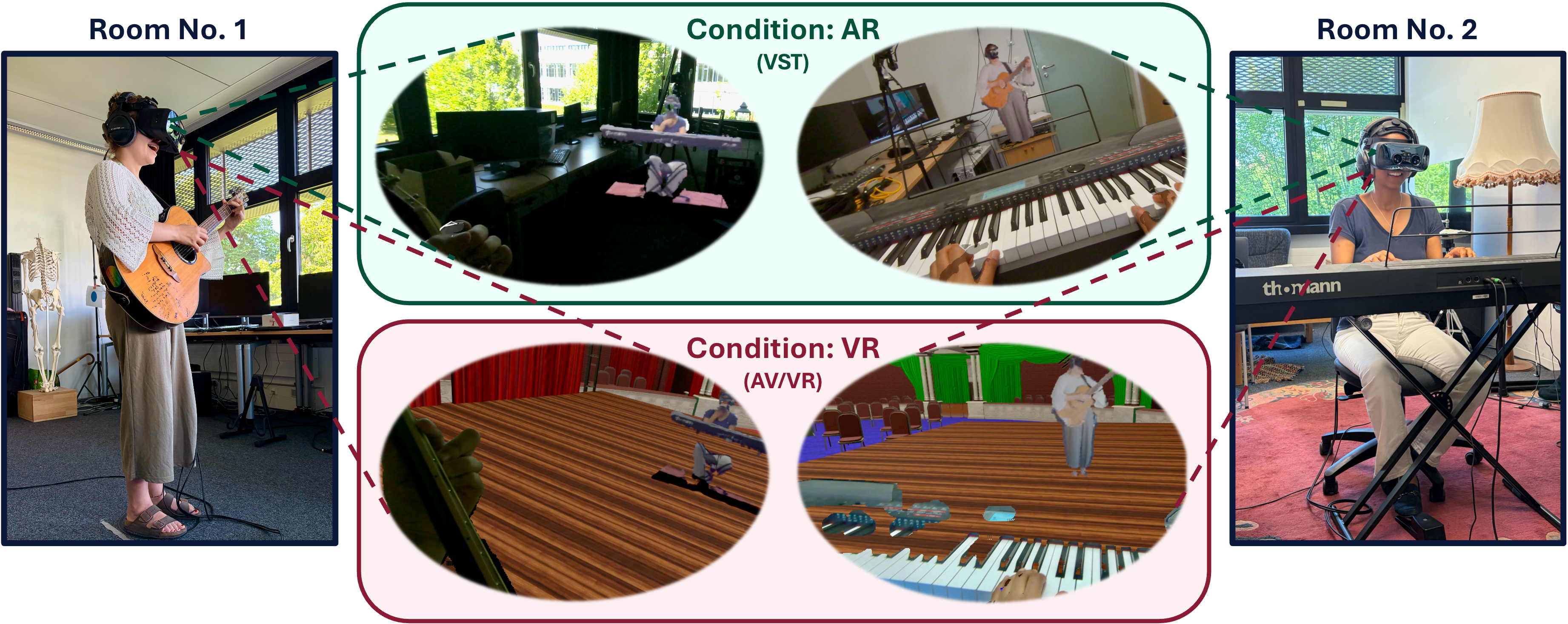}
	\caption{An overview illustration of our approach.}
  	\label{fig:overview}
\end{figure*}

\begin{abstract}

Extended Reality (XR) technologies can place musicians in virtual, mixed, or real environments during networked music performance (NMP).
To examine how NMP applications in XR should be positioned on the reality-virtuality continuum, we conducted a laboratory study (\textit{N} = 28) comparing a simulated NMP experience in Virtual Reality (VR) with one in Augmented Reality (AR).
In both conditions, musicians saw their remote partners in real time as 3D point clouds rendered with an optimized Gaussian Splatting shader.
While VR allowed participants to select and customize fully virtual 3D environments, AR showed a pass-through view of the laboratory environment.
We measured creativity support, flow, involvement, co-presence, and affect, and complemented these data with qualitative interviews on participants' preferences and requirements for creative environments.
Our results show that participants felt significantly more involved by the immersive XR experience in VR than in AR.
Positive affect, flow, and creativity support were strongly correlated in both conditions, whereas involvement correlated with flow and creativity support only in VR.
Qualitative findings showed substantial variation in participants' preferences for AR or VR and also regarding the design of creative virtual environments.
Together, these findings suggest that the surrounding environment is a critical design dimension in networked Musical XR, shaping collaborative musical experiences beyond merely representing remote collaborators and mitigating communication latency. 
Future XR-mediated NMP systems should therefore treat the surrounding space as an integral component of the collaborative musical experience, rather than as a mere visual backdrop. 
At the same time, they should support experiences across a broad spectrum of the reality–virtuality continuum to accommodate highly individual user requirements.
%TODO: remove this before cameraready
%\medskip
%\medskip
%\medskip
\end{abstract}

%\IEEEpeerreviewmaketitle


\section{Introduction}

Extended Reality (XR), an umbrella term encompassing Virtual Reality (VR), Augmented Reality (AR), and Mixed Reality (MR), has rapidly evolved in recent years. 
In particular, advances in MR and video-see-through technologies have significantly enhanced XR's potential for productive and creative work, making music in XR increasingly viable.
Recent work has shown that such applications cannot only provide immersive means for single-user musical creativity \cite{schlagowskiFlowBeatHumanCentered2022,schlagowskiXRCompositionWild2024}, but can also increase the sense of co-presence \cite{bioccaNetworkedMindsMeasure2001}, namely the feeling of ``being there'' with another remote person while making music together \cite{schlagowskiWishYouWere2023,boemImmersionUsabilityComparative2025}. 
This enables a more immersive experience of making music together in real time, commonly referred to as ``jamming'' among musicians, and, when done in a remote setting, falls within the field of Networked Music Performance (NMP). 
Such remote jam sessions may, in some situations, e.g., the recent COVID pandemic, be especially valuable for musicians, as the option of meeting in person may be unavailable. 
Musicians engaged in such real-time jam sessions often improvise, a vital part of musical creativity, which places high demands on audio latency and quality \cite{schlagowskiWishYouWere2023,schlagowskiJammingMRRealTime2022}.

XR-mediated NMP is often discussed in terms of communication latency, packet loss, interaction techniques, and the representation of remote musicians, for example, through volumetric or point-cloud renderings \cite{turchetNetworkedMusicalXR2022}.
However, comparatively little attention has been paid to the surrounding 3D spaces in which musicians collaborate and experience telepresence.
This spatial dimension is particularly relevant in XR, as musicians may either remain situated in their physical surroundings or perform inside fully Virtual Environments (VEs) that can be selected, customized, and potentially shared.

Designing such environments, however, is non-trivial.
In AR, virtual content, such as remote musicians, is embedded into the performer's physical surroundings.
In VR, by contrast, musicians can be placed in familiar musical settings, such as studios or stages, as well as in abstract, natural, or otherwise inaccessible environments, such as spaceships or underwater worlds.
Moreover, VR environments can be experienced either collaboratively, with multiple musicians sharing the same virtual space, or individually, with each musician performing in a different environment while remaining connected through the shared musical interaction.
Consequently, the surrounding space becomes an integral component of the collaborative musical experience rather than merely a visual backdrop.

\begin{figure*}[ht!]
    \centering
    \includegraphics[width=0.32\textwidth]{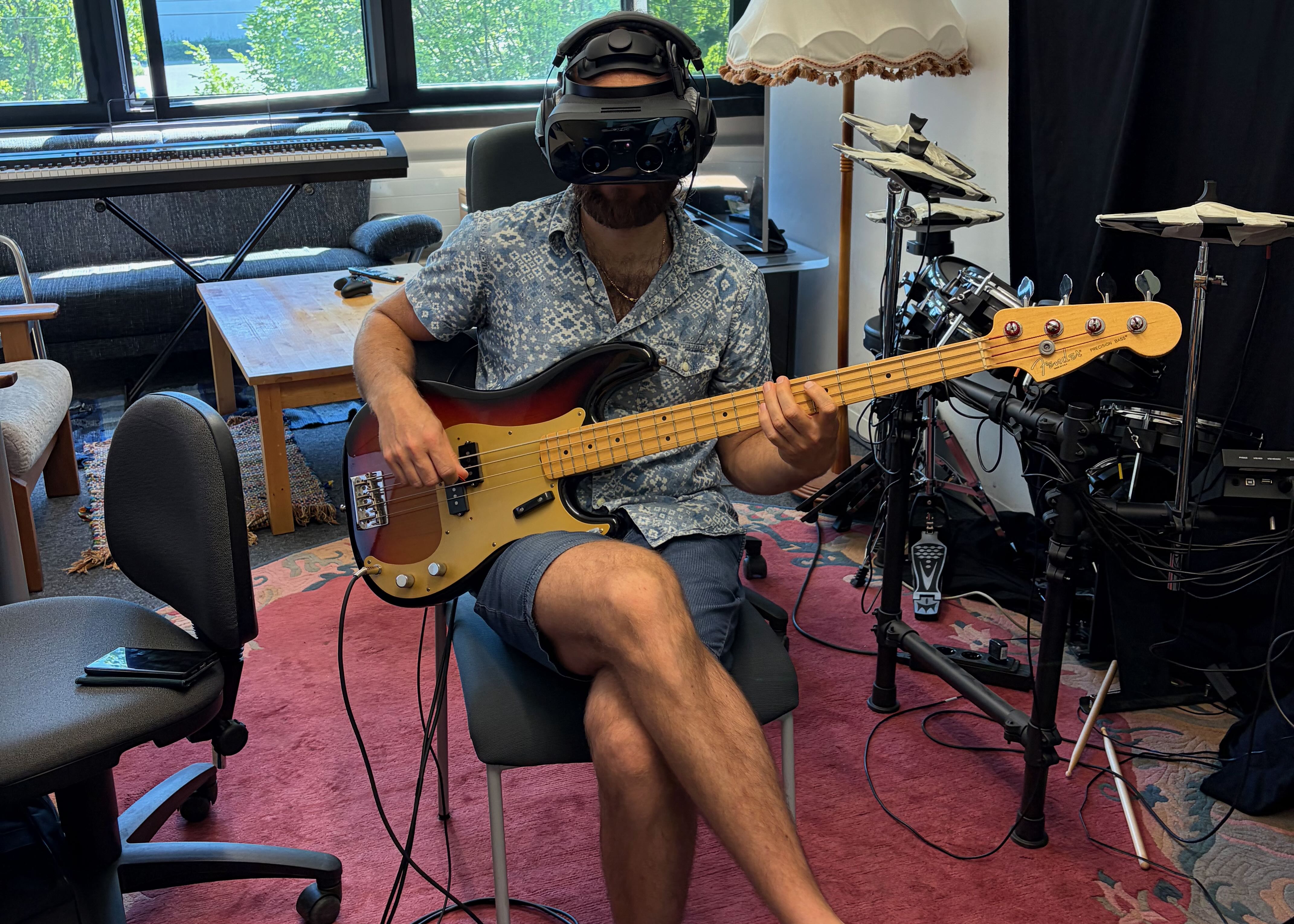}
    \hspace{\figgap}
    \includegraphics[width=0.32\textwidth]{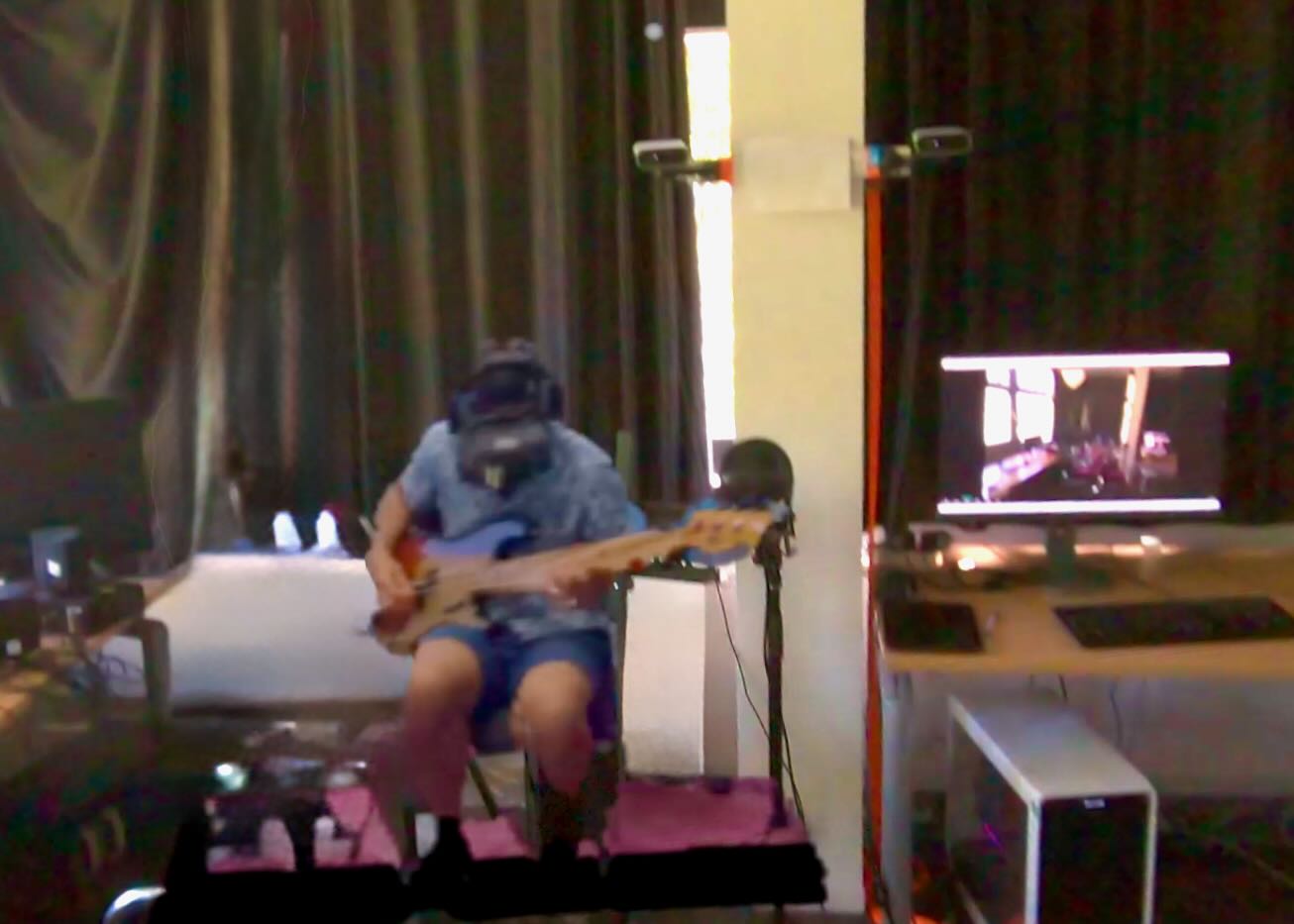}
    \hspace{\figgap}
    \includegraphics[width=0.32\textwidth]{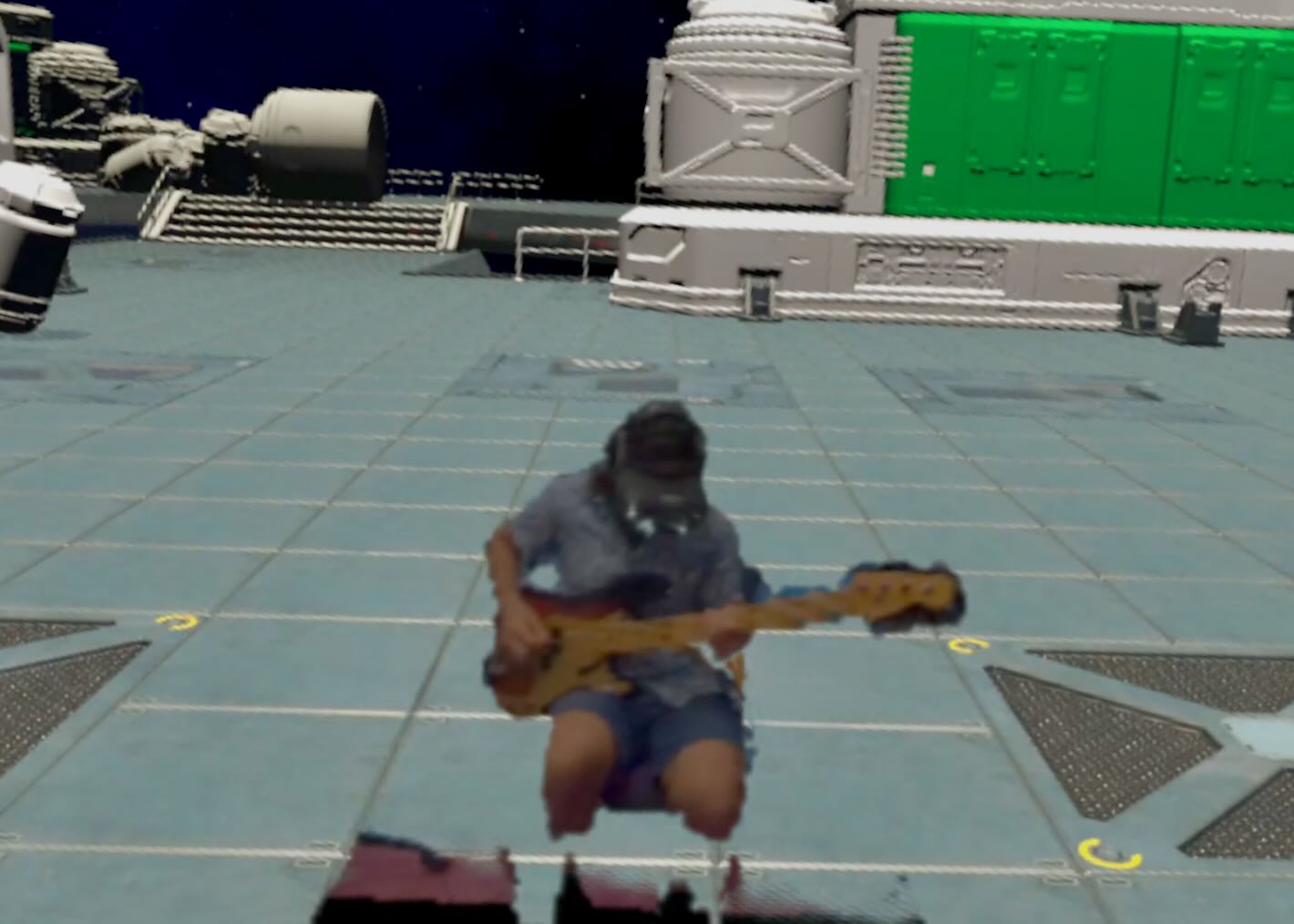}

    \caption{A musician in room 2 (left) and their point cloud in the Point of View (POV) of another musician in room 1 (middle) in the AR condition. The right figure shows their point cloud in the futuristic space environment in the VR condition.}
    \label{fig:pc-examples}
\end{figure*}

To investigate the role of the surrounding space in XR-mediated collaborative music making, we conducted a laboratory study comparing two representative settings using qualitative and quantitative methods.
In the \textit{AR condition}, musicians remained situated in their physical laboratory surroundings while interacting through pass-through video and 3D point-cloud representations of their remote partners.
In the \textit{VR condition}, musicians performed in customizable virtual spaces, with the option to share the same space or independently select different ones, whilst seeing the same point cloud as in AR.
By comparing these two settings (see \autoref{fig:pc-examples} for more detailed POV screenshots), we investigate how the design of the surrounding space influences creativity support, affect, co-presence, involvement/presence, and flow during remote musical improvisation.

By analyzing data from self-report questionnaires, semi-structured interviews, and objective logs, we approached the following four central research questions of our study:
\begin{itemize}
    \item \textit{RQ1:} How does providing musicians with immersive, customizable VEs in the VR condition influence creativity support \cite{cherryQuantifyingCreativitySupport2014}, affect \cite{watsonDevelopmentValidationBrief1988a}, co-presence \cite{bioccaNetworkedMindsMeasure2001}, involvement (a scale related to presence \cite{steuerDefiningVirtualReality1995}), and flow \cite{csikszentmihalyiFlowPsychologyOptimal2008} compared to performing within their real surroundings using AR?
    \item \textit{RQ2:} Can we observe correlations between psychometrics and the time musicians spent making music together in similar or different VEs within the VR condition?
    \item \textit{RQ3:} Do musicians prefer jamming in the AR condition or in the VR condition, and how do they explain their preference qualitatively?
    \item \textit{RQ4:} What are musicians' requirements for creative environments, and does our VR apparatus featuring a 3D User Interface that allows for VE customization meet these requirements?
\end{itemize}

\section{System Implementation}

\subsection{Apparatus}

In our laboratory study setup, we connected two separate rooms (see \autoref{fig:overview}) within the same building to simulate an NMP scenario in XR for diads under ideal network conditions (i.e., minimal latency and packet loss). 
The technical setup consisted of two Varjo XR-3 HMDs\footnote{\url{https://varjo.com/products/varjo-xr-3}} connected to two high-performance PCs, each featuring an Intel Core i9-13900K CPU, an Nvidia RTX 4090 GPU, and 128 GB of RAM, one in each lab, each running our Unity\footnote{\url{https://unity.com}} projects for the XR application. 
Each computer was connected to two Azure Kinect DK Cameras\footnote{\url{https://azure.microsoft.com/de-de/products/kinect-dk}} to capture 3D point clouds at a higher resolution than is typically possible with a single camera. 

To avoid confounding variables related to audio latency or quality, we managed the audio mixes for both jamming partners using a traditional analog mixer connected to bodypacks that drive Beyerdynamic DT 770 Pro studio headphones. 
We connected not only the musicians' instruments to the mixer but also one Rode NT-1A microphone per room to capture room audio, including participants' voices for communication.
The test supervisor responsible for the sound check before jamming sessions communicated with participants via another audio talkback microphone on the mixing console in a separate control room.

\subsection{3D Point Clouds}
\begin{figure}[h!]
    %\centering
    \includegraphics[width=0.361\linewidth]{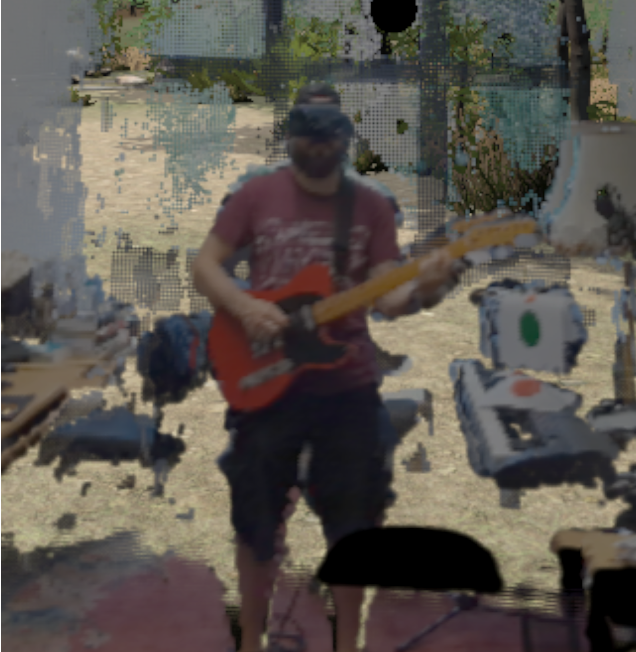}
    \hspace{0.01\linewidth}
    \includegraphics[width=0.361\linewidth]{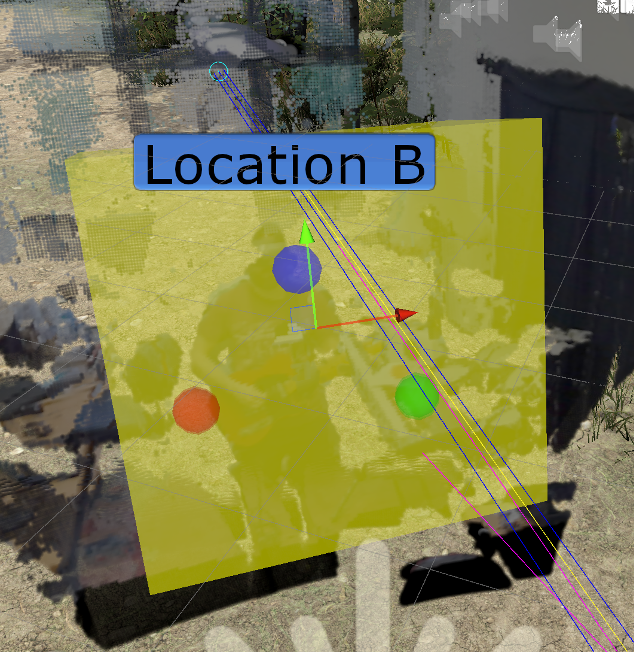}
    \hspace{0.01\linewidth}
    \includegraphics[width=0.22\linewidth]{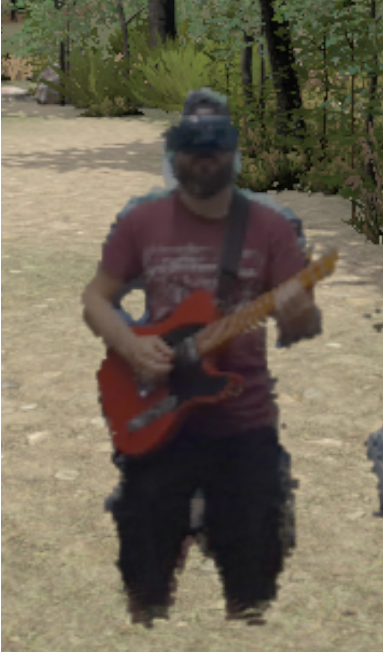}

    \caption{Masking of the point cloud using a bounding box. Left is the unfiltered point cloud, in the middle the bounding box in the Unity Editor, and on the right the masked point cloud, which is sent via UDP.}
    \label{fig:filter}
\end{figure}

To optimize latency, 3D point clouds were filtered with a mask before transmission to remove unnecessary data before transmission. 
We achieved this with a custom-built data-processing pipeline implemented in the Social Signal Interpretation (SSI) framework \cite{wagnerSocialSignalInterpretation2013a} and custom prefabs in Unity, which could be used to set filter parameters. 
The mask was applied visually using a bounding box, which could be oriented and scaled in the Unity Editor (see \autoref{fig:filter}).

To properly align the point clouds from both depth cameras, we calibrated the filtered 3D point clouds before each study iteration. 
To this end, we implemented a procedural calibration that automatically aligns distinct 3D targets (spheres/circles) by iteratively translating, rotating, and scaling the respective point clouds until they match.
To set a target for the procedural alignment algorithm, the test supervisor would place three colored spheres (visible in the middle screenshot of \autoref{fig:filter}) on top of a real-world human-sized calibration model while viewing its point cloud in the Unity Editor.

\begin{figure}[hb!]
    %\centering
    \includegraphics[width=0.481\linewidth]{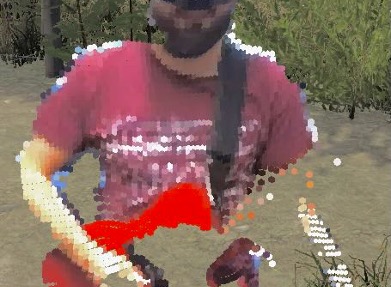}
	\hspace{0.001\textwidth}
	\includegraphics[width=0.481\linewidth]{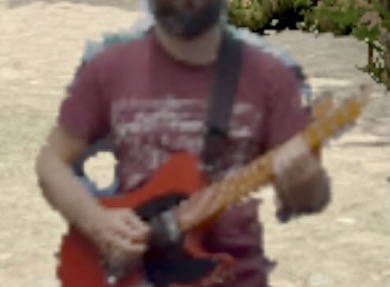}

    \caption{A point cloud using a camera-oriented disc shader (left), and one using our Gaussian splatting shader (right).}
    \label{fig:shader}
\end{figure}

For rendering the 3D point clouds, we used a shader based on Gaussian Splatting \cite{kerbl3dGaussianSplatting2023}, which blended color data with a Gaussian distribution to achieve a more seamless transition between rendered points. 
Compared to the shader used in previous work \cite{schlagowskiWishYouWere2023}, this increased the level of detail in the point clouds while also improving accidental false masking of darker regions of point clouds (see \autoref{fig:shader}).
%These blended 3D point clouds were rendered in real-time using Unity Engine after being transmitted over the local network via UDP.

Using the Varjo Unity XR SDK\footnote{\url{https://developer.varjo.com/docs/unity-xr-sdk/unity-xr-sdk}}, we keyed out objects close to the camera with the depth occlusion functions\footnote{\url{https://developer.varjo.com/docs/unity-xr-sdk/depth-occlusion-with-varjo-xr-plugin}}, so that musicians could see their instruments and hands while making music not only in pass-through AR, but also in VR.
Using a local broadband network to send our filtered point cloud data via UDP and by using high-performance PCs, we reduced point cloud latency to below a noticeable threshold relative to audio latency.

%\begin{figure*}[ht!]
%    \centering
%    \includegraphics[width=0.32\textwidth]{filt-01.png}
%    \hspace{0.01\textwidth}
%    \includegraphics[width=0.32\textwidth]{filt-02.png}
%    \hspace{0.01\textwidth}
%    \includegraphics[width=0.193\textwidth]{filt-03.png}
%
%    \caption{Filtering of the point cloud.}
%    \label{fig:filter}
%\end{figure*}

%\begin{figure}[h!]
%    \centering
%    \includegraphics[width=\linewidth]{comparison-gaussian.png}
%    \caption{3D point clouds of a musician featuring camera-oriented discs (left) and Gaussian Splatting (right).}
%    \label{fig:result-fks}
%\end{figure}

\begin{figure*}[ht!]
\centering
\begin{minipage}{0.495\textwidth}
  \centering
  \includegraphics[width=\linewidth]{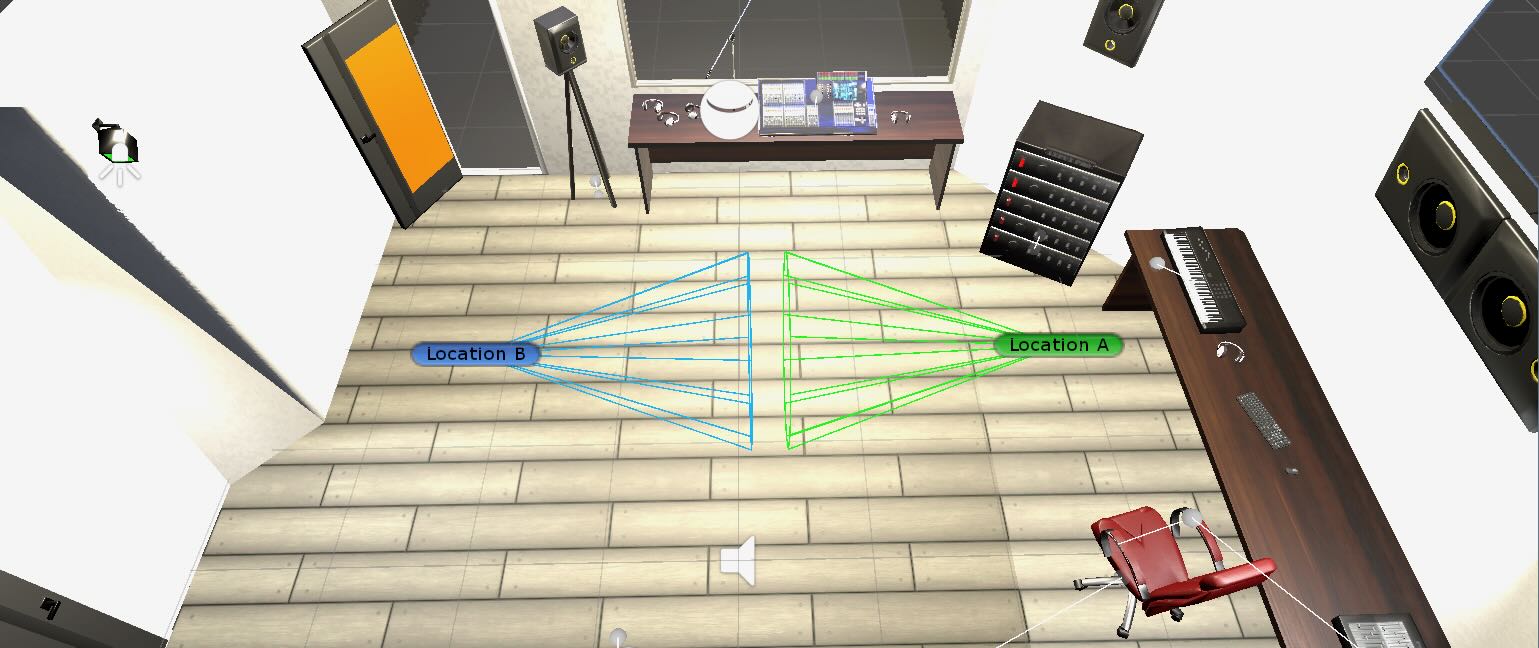}
\end{minipage}\hfill
\begin{minipage}{0.495\textwidth}
  \centering
  \includegraphics[width=\linewidth]{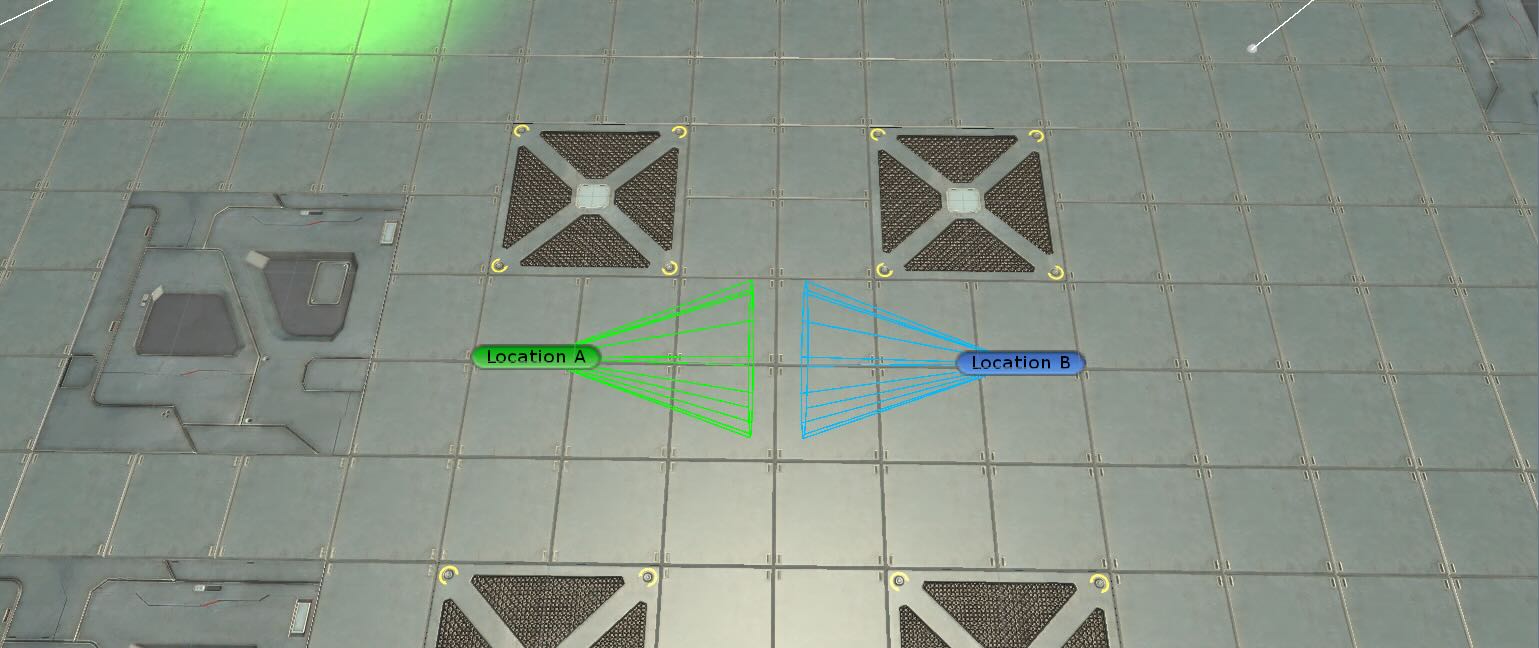}
\end{minipage}

\vspace{0.5	em}

\begin{minipage}{0.495\textwidth}
  \centering
  \includegraphics[width=\linewidth]{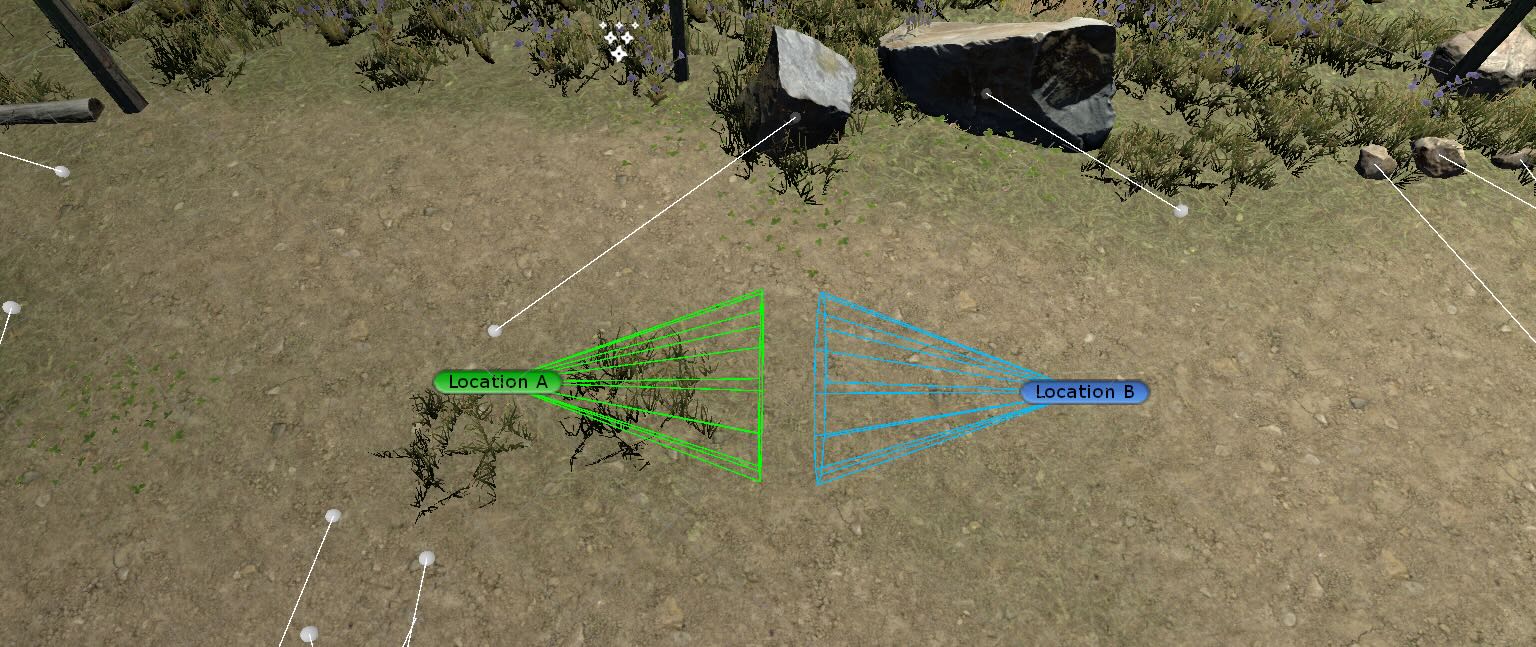}
\end{minipage}\hfill
\begin{minipage}{0.495\textwidth}
  \centering
  \includegraphics[width=\linewidth]{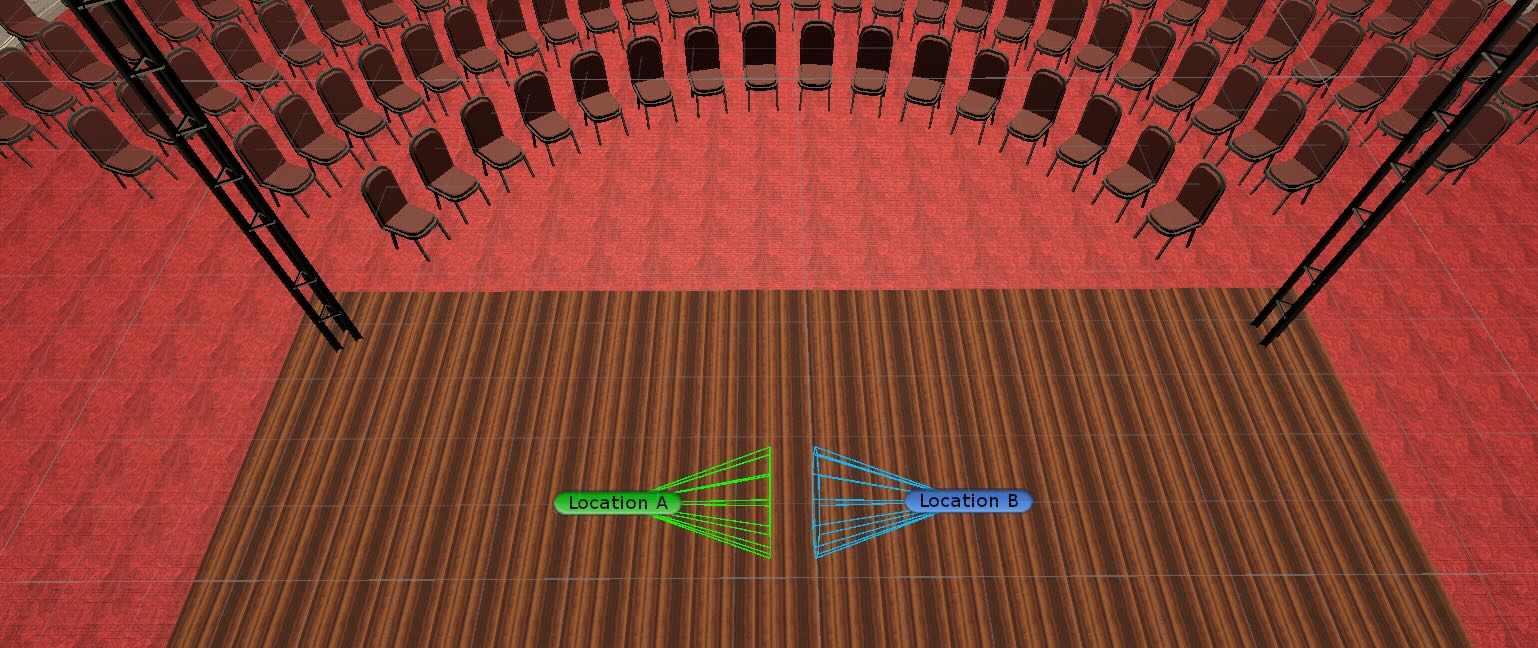}
\end{minipage}
\caption{The placements of both musicians within four of the selectable and customizable creative virtual environments (studio, space, meadow, and theater) when they were present in the same environment.}
\label{fig:environments-placements}
\end{figure*}

\subsection{Customizable Virtual Environments}

In order to provide musicians with means to select and customize their environments according to their individual needs, we implemented a 3D user interface using the MRTK framework\footnote{\url{https://learn.microsoft.com/en-us/windows/mixed-reality/mrtk-unity/mrtk3-overview/}} where they could select among the four environments depicted in \autoref{fig:environments-placements}, which we adapted from a previous study on creative environments \cite{schlagowskiFlowBeatHumanCentered2022}. 
The four general Virtual Environments (VEs) (see \autoref{fig:environments-placements}), which we adapted, were:
\begin{enumerate}
    \item \textit{Studio:} A recording studio environment featuring instruments and a performance and mixing room. There was a brightly lit and a dimmed version available;
    \item \textit{Space:} A futuristic environment featuring a floating platform and a space skybox;
    \item \textit{Meadow:} A natural environment featuring a green meadow;
    \item \textit{Theater:} A theater environment with a stage in front of seats (no audience present).
\end{enumerate}

For each of these VEs, certain parameters were customizable to better suit the musicians' individual needs. 
There were three groups of adjustable parameters, originally derived from a literature survey on environmental features which could be creatively stimulating, reported in \cite{schlagowskiFlowBeatHumanCentered2022}:
\begin{enumerate}
    \item \textit{Visual Complexity:} Number of objects in the user environment and how chaotically they are arranged;
    \item \textit{Lighting:} Time of day, color temperature of the main light, number of accent lights, color of accent lights;
    \item \textit{Color:} Number of colored objects and the color of these objects.
\end{enumerate}

Participants could customize these parameters in the 3D user interface using the hand-tracking capabilities provided by the Varjo SDK. 
When finished, they could close the UI to have a full view of their customized VEs while jamming. 
Participants were free to open the menu at any time when looking at their inner hand and pushing a button that then appeared.
If participants chose the same environment type (studio, space, meadow, or theater), they were placed in distinct positions within that environment, facing each other (see the red and green placements in \autoref{fig:environments-placements}).

% \begin{figure}[ht!]
% 	\centering
% 	\includegraphics[width=.99\linewidth]{ar-pointcloud.jpeg}
% 	\caption{A POV frame of a remote musician playing bass rendered in the AR condition using the Gaussian Splatting shader and video see-through.}
%   	\label{fig:ar-pointcloud}
% \end{figure}

% \begin{figure}[ht!]
% 	\centering
% 	\includegraphics[width=.99\linewidth]{vr-pointcloud.jpeg}
% 	\caption{A POV frame of the same point cloud as in \autoref{fig:ar-pointcloud} rendered in the VR condition in a futuristic environment.}
%   	\label{fig:vr-pointcloud}
% \end{figure}

% Preamble:
% \usepackage{graphicx}

\begin{figure*}[ht!]
	\centering
	\includegraphics[width=.7\textwidth]{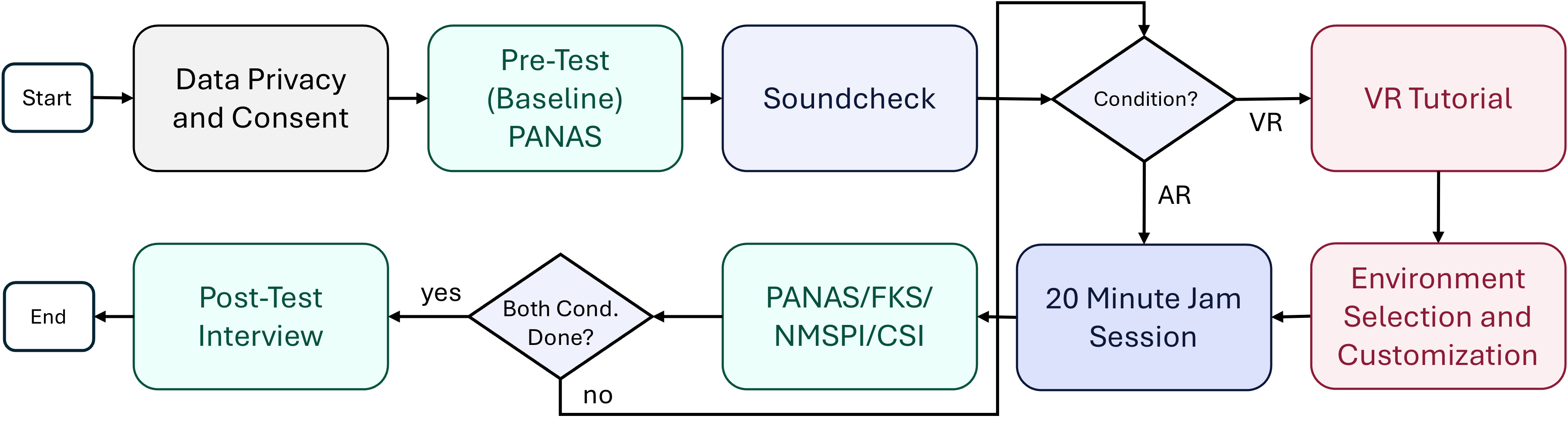}
	\caption{Study Procedure.}
  	\label{fig:procedure}
\end{figure*}

\section{User Study}

To address research questions RQ1-RQ4, we conducted a laboratory study in Germany between April and June 2026 that employed both quantitative and qualitative methods. 

\subsection{Participants}

We recruited a total of 28 participants (two more than required as calculated in an a priori power analysis) via mail, social media posts, printed posters, and contact with local jamming ensembles. 
Of these, 22 were males, 5 females, and 1 non-binary, with a mean age of 30.3 (SD = 8.1). 
Five participants assessed their skill as beginner level, 18 as intermediate, and five as professional.
Nine of the participants played the guitar, five played the drums, nine played the piano, three played bass, one played saxophone, one reported singing/vocals, and one played the flute.
We encouraged participants to bring jamming partners they liked and also bring musical instruments of their choice. 
If they did not have jamming partners available, we offered to pair them with other musicians who matched their skill levels. Also, we provided electronic drums, an electric guitar, and a stage piano on-site in case they needed them.
The experiment took an average of two hours. Each participant received a total of 30€ in compensation.

\subsection{Study Procedure}

Our laboratory study was structured as a \textit{within-subjects} experiment in which each participant experienced both conditions (AR and VR) in a randomized order, with questionnaires after each session and a post-test interview at the very end. The study procedure is depicted in \autoref{fig:procedure} and took about two hours in total.
After Data and Privacy Consent, participants completed a Pre-Test PANAS questionnaire \cite{watsonDevelopmentValidationBrief1988a} to assess their affect before jamming in either condition and to put subsequent PANAS measurements into context.
Then, they proceeded to the soundcheck, where the test supervisor ensured that each participant was satisfied with their individual blend of room and instrument audio from themselves and their jamming partners.
If participants were in the VR condition, they received an additional tutorial and instructions on using hand tracking and the 3D user interface to select and customize their VEs. 
Furthermore, they were told that they were free to make these choices either for themselves or in conversations with their musical partners, and that they could also change their environment during jam sessions if desired.
As soon as both the audio and XR setups were properly configured, the test supervisors told participants they had 20 minutes to play and encouraged them to improvise rather than play songs they already knew.
After each jam session, participants completed our self-assessment instruments measuring the dependent variables for research questions RQ1 and RQ2. 
After the two jam sessions and both rounds of questionnaires were completed, participants were interviewed for about 5-10 minutes to gather qualitative data for research questions RQ3 and RQ4.

\subsection{Dependent Variables and Hypotheses}

%TODO: Messgrößen evtl. nochmal erklären  
% co-presence: the ``degree to which the observer believes he/she is not alone and secluded, their level of peripherally or focally awareness of the other, and their sense of the degree to which the other is peripherally or focally aware of them'' \cite{biocca_networked_2001} 
The dependent variables and questionnaires we used to measure them are listed in \autoref{table:variables}.
For all questionnaires, we used German versions. 
However, while the German versions of PANAS, IPQ, and FKS were psychometrically validated, the German NMSPI version was adapted from a previous study \cite{schlagowskiWishYouWere2023}, in which it was employed successfully.
For \textit{VE Overlap}, we used logs of participants' use of the 3DUI for environment selection and customization to calculate the ratio of time spent in the same Virtual Environment (VE) to total time spent in the VR condition.
Based on prior studies and the overall research questions, we formulated seven hypotheses before conducting the experiment:
%First, because environment customization was only available in VR, we hypothesized that the CSI rating would be higher in VR than in AR (H1).
%Also, observing an increase in the ``wow'' factor in VR led us to assume an increase in positive affect, as measured by PANAS, in the VR condition compared to AR (H2).
%Furthermore, we expected the involvement subscale of the IPQ, which describes the relative attention to virtual versus real environments \cite{schubertExperiencePresenceFactor2001b}, to be higher in VR (H3).
%Additionally, since we used the same means of showing jamming partners in AR and VR (3D point clouds with Gaussian splatting) across both conditions, we hypothesized that co-presence would be equal between conditions (H4).
%Our fifth hypothesis (H5) states that there is no difference in flow between conditions, as previous work has suggested that flow, which can make musicians potentially unaware of their surroundings, may limit any impact of user surroundings \cite{schlagowskiXRCompositionWild2024,schlagowskiMusicalCreativityFlow2026}.
%For a better overview, we list all seven hypotheses below. 
%We report on the statistical tests for them in \autoref{sec:results}.
\begin{itemize}
    \item [H1:] Creativity support index (CSI) scores are higher in the \emph{VR} condition than in the \textit{AR} condition, as our customizable immersive virtual environments may provide creativity-supportive affordances by allowing participants to actively shape task-relevant contextual cues, experiential framing, and engagement \cite{bourgeois2022immersive}.
    \item [H2:] PANAS scores for positive affect are higher in the \emph{VR} condition compared to the \textit{AR} condition, as VR can elicit awe-like affective responses, particularly through immersive and presence-enhancing environments \cite{chirico2018designing}.
    \item [H3:] Scores on the involvement sub-scale of the IPQ are higher in the \emph{VR} condition compared to the \textit{AR} condition, reflecting the expectation that selectable virtual surroundings increase users' attentional engagement with the mediated environment \cite{bourgeois2022immersive,schubertExperiencePresenceFactor2001}.
	\item [H4:] Mutual awareness/co-presence scores, as measured by the NMSPI, do not differ between the \emph{AR} and \emph{VR} conditions (as point clouds do not differ between conditions).
	\item [H5:] Flow scores, as measured by the FKS, do not differ between the \emph{AR} and \emph{VR} conditions. This would be in line with a central hypothesis in \cite{schlagowskiMusicalCreativityFlow2026}.
    \item [H6:] There is a positive correlation between the VE Overlap scores and the involvement scores of the IPQ, as a spatially situated and dialogically consistent interaction partner may increase the coherence of the virtual experience and draw users’ attention more strongly into the virtual environment \cite{schubertExperiencePresenceFactor2001}.
    \item [H7:] There is a positive correlation between the VE Overlap score and the mutual awareness/co-presence scores (NMSPI), as co-location may provide additional non-mediated social and spatial cues that increase mutual awareness and perceived social presence \cite{biocca2003toward}.
\end{itemize}

% \begin{figure*}[ht!]
%     \centering
%     \includegraphics[width=0.9\textwidth]{overview-plot.png}
%     \caption{A normalized plot of all measured dependent variables assigned to hypotheses H1-H7.}
%     \label{fig:overview-plot}
% \end{figure*}

\begin{table}[h!]
    \caption{An overview of dependent variables, their respective measuring tools, and hypotheses.}
    \begin{tabular}{p{0.43cm}|p{1.35cm}|p{3.07cm}|p{0.8cm}|p{0.9cm}}
    \toprule
    \textbf{RQ} & \textbf{Variable} & \textbf{Measured with} & \textbf{Range} & \textbf{Hypoth.}\\ 
    \midrule

    1, 2 & Creativity Support &  Creativity Support Index (CSI) \cite{cherryQuantifyingCreativitySupport2014} & 0-100 & H1 \\

    \rowcolor{Grey}
    1, 2 & Positive Affect &  Positive and Negative Affect Schedule (PANAS) \cite{watsonDevelopmentValidationBrief1988a} & 10-50/ 1-5 & H2\\

    1, 2 & Involvement &  IGroup Presence Questionnaire (IPQ) \cite{schubertExperiencePresenceFactor2001} & 1-7 & H3, H6 \\

    \rowcolor{Grey}
    1, 2 & Co-presence & Networked Minds Social Presence Inventory (NMSPI) \cite{bioccaNetworkedMindsMeasure2001} & 1-7 & H4, H7 \\
    
    1, 2 & Flow Level & Flow Kurzskala (FKS) \cite{rheinbergErfassungFlowErlebens2003b} & 1-7 & H5 \\
    
    \rowcolor{Grey}
    2 & VE Overlap & Logs/objective data & 0-1 & H6, H7 \\
    
    \bottomrule
    
    \end{tabular}

    \label{table:variables}
\end{table}

\begin{table*}[]
	\tiny
	\centering
	\caption {Overview of the results of the confirmatory hypothesis tests} 
	\label{tab:merged_results}
	\resizebox{\linewidth}{!}{
		\begin{tabular}{llcccccc}
			\toprule
			\textbf{Condition}
			& 
			& \textbf{AR}
			& \textbf{VR}
			&
			&
			&
			\\
			\rowcolor{Grey}
			\textbf{Hypothesis}
			& \textbf{Test type}
			& \textit{M(SD)}
			& \textit{M(SD)}
			& \textbf{Statistic}
			& \textbf{$r$}
			& \textbf{$p$}
			& \textbf{corrected $p$} \\
			\midrule
			
			H1 
			& One-tailed \textit{t}-test 
			& 61.10 (16.56) 
			& 61.15 (17.68) 
			& \textit{t}(26) = -0.02 
			& .00 
			& .49
			&  1.0\\
			
			\rowcolor{Grey}
			H2 
			& One-tailed \textit{t}-test 
			& 3.50 (0.63) 
			& 3.52 (0.68) 
			& \textit{t}(27) = -0.14 
			& .03 
			& .44
			&  1.0\\
			
			\textbf{H3} 
			& \textbf{One-tailed \textit{t}-test} 
			& \textbf{3.55 (0.85)} 
			& \textbf{4.93 (1.19)} 
			& \textbf{\textit{t}(27) = -5.12} 
			& \textbf{.70} 
			& \textbf{\textless.001***}
			& \textbf{\textless.001***} \\
			
			\rowcolor{Grey}
			H4 
			& TOST \textit{t}-test 
			& 4.12 (0.36) 
			& 3.97 (0.30) 
			& $\delta = 0.244$ 
			& 
			& .15
			& .76\\
			
			H5 
			& TOST \textit{t}-test 
			& 4.95 (0.99) 
			& 4.74 (1.26) 
			& $\delta = 0.429$ 
			& 
			& .09
			&  .55\\
			
			\rowcolor{Grey}
			H6 
			& Correlation, Pearson 
			& 
			& 
			& 
			& -.12 
			& .73
			&  1.0\\
			
			H7 
			& Correlation, Pearson 
			& 
			& 
			& 
			& -.00 
			& .51
			&  1.0\\
			
			\bottomrule
			***\textit{p} \textless .0001
		\end{tabular}
	}
\end{table*}

\subsection{Confirmatory Hypothesis Tests}
We tested hypotheses H1-H3 using t-tests, equivalence tests for H4 and H5, and correlation analysis for H6 and H7.
Prior to t-tests, assumptions of normality were evaluated using the Shapiro-Wilk test \cite{shapiroAnalysisVarianceTest1965a}. 
After removing one outlier (a participant who selected only maximum values on the CSI in one condition), both assumptions were met across all the data we tested. 
Consequently, we use two-tailed paired-samples \textit{t}-tests \cite{studentProbableErrorMean1908a} for comparisons.
For the equivalence hypotheses H4 and H5, two one-sided tests were conducted using TOST \cite{schuirmannComparisonTwoOnesided1987} with an equivalence bound of $\Delta = 0.5$. 
Associations related to H6 and H7 were examined using Pearson's correlation coefficients \cite{pearsonVIINoteRegression1895} as the data were metric and showed linear trends upon visual inspection. 
To control for multiple comparisons, \textit{p}-values were adjusted using the Holm-Bonferroni method \cite{holmSimpleSequentiallyRejective1979a}.

\subsection{Exploratory Comparisons and Correlation Analysis}
In addition to the confirmatory hypothesis tests, we conducted exploratory comparisons between the AR and VR conditions for all remaining  PANAS, NMSPI, FKS, and IPQ subscales. 
These analyses were not associated with predefined hypotheses and should therefore be interpreted as exploratory. 
For each comparison, normality was assessed using the Shapiro-Wilk test, and equality of variances was evaluated using Levene's test, as in testing H1-H3. 
If both assumptions were met, differences between conditions were examined using two-tailed paired-samples \textit{t}-tests. 
If either assumption was violated, the nonparametric Wilcoxon signed-rank test \cite{wilcoxonIndividualComparisonsRanking1992} was used instead. Here, this was the case for three tests.
No correction for multiple comparisons was applied to these exploratory analyses.
We report on these tests in \autoref{tab:stat_comparison}. 

%\subsection{Exploratory Correlation Analysis}
Furthermore, we conducted an exploratory correlation analysis to examine associations among flow, copresence, involvement, and positive affect, separately for each condition (AR and VR). 
Pearson correlation coefficients were used because the variables were treated as continuous and visual inspection indicated approximately linear relationships.
For VR, we additionally calculated correlations with VE Overlap. 
We report on this analysis in \autoref{subsec:corr-analysis}.

\subsection{Interview and Qualitative Analysis}
\label{subsec:quali-methods}

The final step of our study procedure was a short, semi-structured interview with each participant, lasting 5-10 minutes. 
This interview was deliberately designed to obtain qualitative insights contributing to research questions RQ3 and RQ4.
We also posed additional questions on general feedback. 
However, due to the page limit of this paper, we did not report on these questions here.
The questions we asked participants (translated from German) were:
\begin{enumerate}
    \item Which was better, AR or VR, and why?
    \item What aspects of an environment are important to you for being musically creative?
    \item Which environmental settings did you adjust, and why?
    %\item In which session did you find it easier to be musically creative?
    %\item Would you use this system if it were released as a product? Why or why not?
    %\item Do you have any additional feedback on the AR experience?
    %\item Do you have any additional feedback on the VR experience?
    %\item Do you have any additional feedback on the study or anything else you would like to mention?
\end{enumerate}

Each interview was recorded on audio and transcribed using whisper-large-v3-german as a local speech-to-text model,\footnote{\url{https://huggingface.co/primeline/whisper-large-v3-german}} and finally analyzed using inductive thematic analysis \cite{braunThematicAnalysis2012a} using the MaxQDA software.\footnote{\url{https://www.maxqda.com/}} 
Here, two authors iteratively coded transcript segments (codes representing themes or general aspects uttered by participants), compared cases, refined categories, and developed an interpretation grounded in the evidence.
For visual representation, we generated word clouds of the most common codes (hereafter referred to as \textit{code clouds}) we found in the interview transcripts (see, e.g., \autoref{fig:ar-preference-reasons}). 
Bigger codes were coded more frequently than smaller codes. 
We report on the results of this analysis in \autoref{subsec:quali-analysis}. 
To make the results more compact, we excluded themes/codes that were uttered just once.

\section{Results}
\label{sec:results}

\begin{figure*}[h!]
    \centering

    \begin{minipage}[t]{0.48\linewidth}
        \centering
        \includegraphics[width=\linewidth]{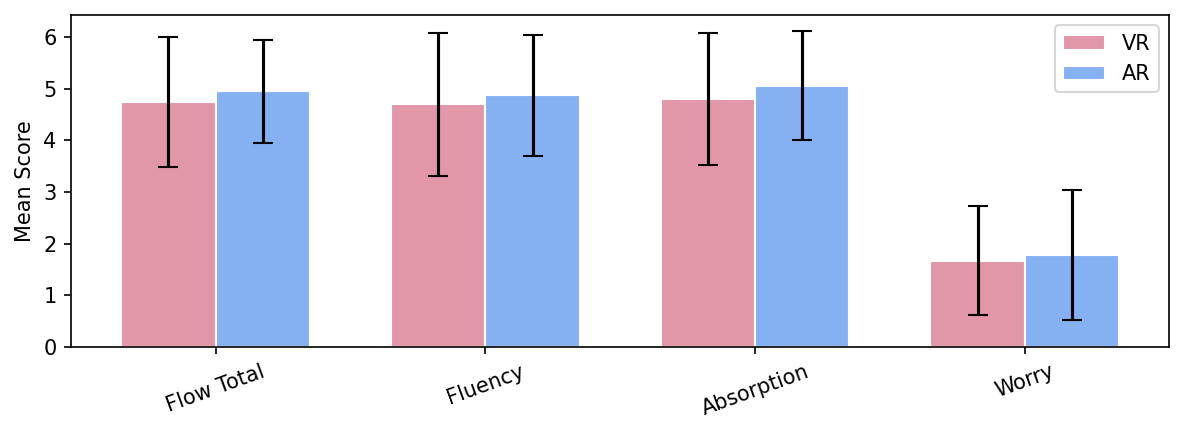}
        \caption{Subscales of the FKS for both conditions.}
        \label{fig:result-fks}
    \end{minipage}
    \hfill
    \begin{minipage}[t]{0.48\linewidth}
        \centering
        \includegraphics[width=\linewidth]{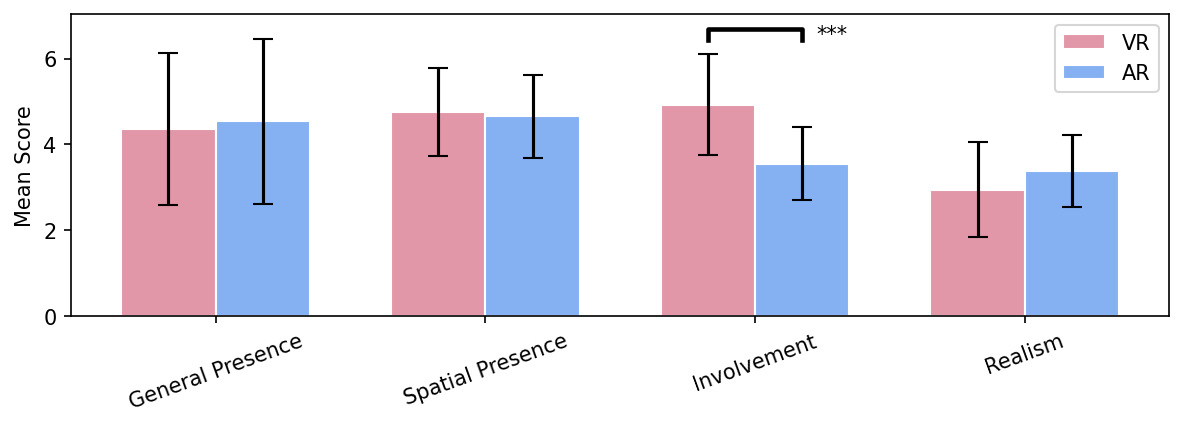}
        \caption{Subscales of the IPQ for both conditions.}
        \label{fig:result-ipq}
    \end{minipage}

    \vspace{0.53cm}

    \begin{minipage}[t]{0.48\linewidth}
        \centering
        \includegraphics[width=\linewidth]{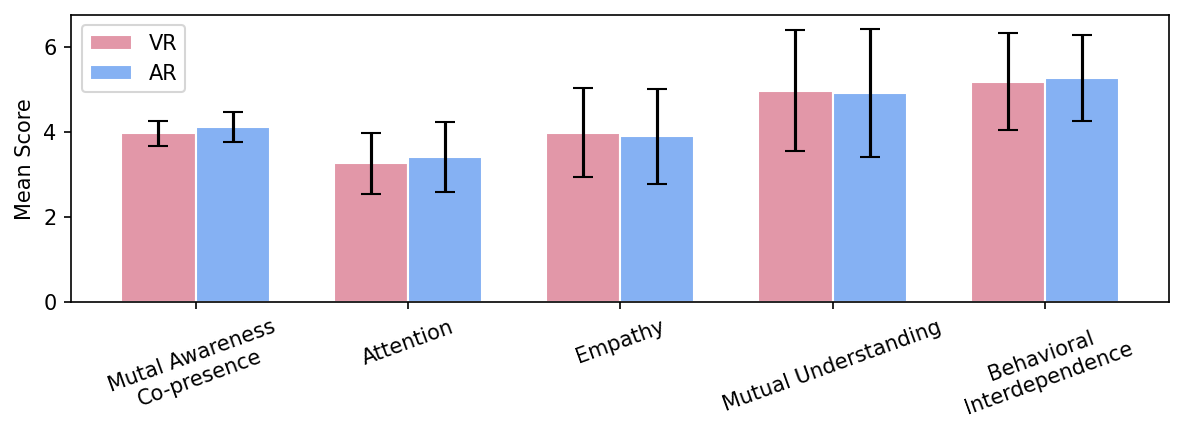}
        \caption{Subscales of the NMSPI for both conditions.}
        \label{fig:result-nmpsi}
    \end{minipage}
    \hfill
    \begin{minipage}[t]{0.48\linewidth}
        \centering
        \includegraphics[width=\linewidth]{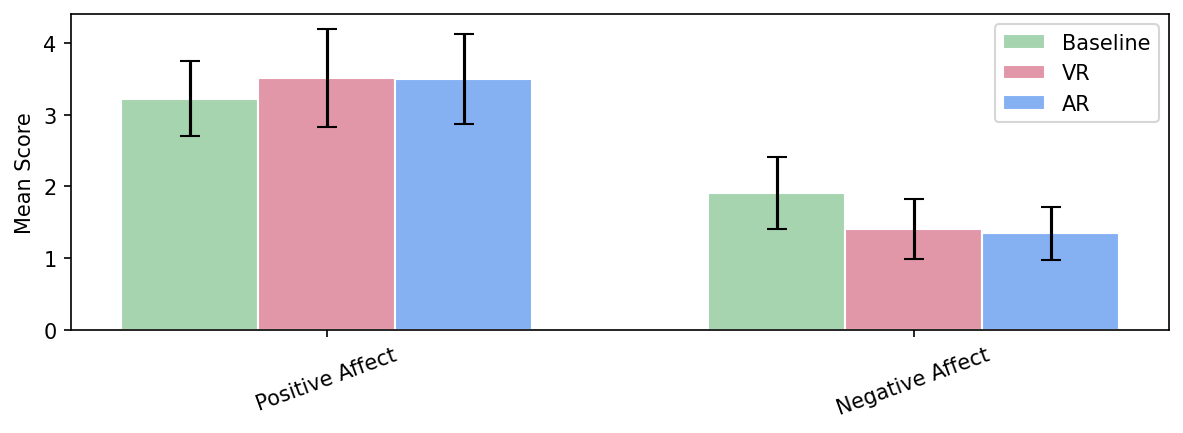}
        \caption{Subscales of the PANAS for both conditions and baseline.}
        \label{fig:result-panas}
    \end{minipage}

    %\caption{Subscales of all questionnaires for both conditions.}
    %\label{fig:result-questionnaires}
\end{figure*}

\subsection{Confirmatory Hypothesis Tests}
\label{subsec:hypoth-tests}

\autoref{tab:merged_results} shows an overview of the results of the confirmatory hypothesis tests.
For \textbf{H1}, there was no significant difference between AR (\textit{M} = 61.10, \textit{SD} = 16.56) and VR (\textit{M} = 61.15, \textit{SD} = 17.68), \textit{t}(26) = -0.02, \textit{r} = .00, \textit{p} = .49, Holm-Bonferroni corrected \textit{p} = 1.00. Thus, H1 was \textit{not supported}.
For \textbf{H2}, the results likewise showed no significant difference between AR (\textit{M} = 3.50, \textit{SD} = 0.63) and VR (\textit{M} = 3.52, \textit{SD} = 0.68), \textit{t}(27) = -0.14, \textit{r} = .03, \textit{p} = .44, corrected \textit{p} = 1.00. Therefore, H2 was \textit{not supported}.
For \textbf{H3}, a significant difference was observed between AR (\textit{M} = 3.55, \textit{SD} = 0.85) and VR (\textit{M} = 4.93, \textit{SD} = 1.19), \textit{t}(27) = -5.12, \textit{r} = .70, \textit{p} $< .001$, corrected \textit{p} $< .001$. The effect was large, indicating that scores were significantly higher in VR than in AR. Thus, H3 was \textit{supported}.
For \textbf{H4}, equivalence between AR (\textit{M} = 4.12, \textit{SD} = 0.36) and VR (\textit{M} = 3.97, \textit{SD} = 0.30) was tested using TOST with $\Delta = 0.5$. The observed effect was $\delta = 0.244$. The equivalence test was not significant after correction, \textit{p} = .15, corrected \textit{p} = .76. Accordingly, equivalence could not be concluded, and H4 was \textit{not supported}.
For \textbf{H5}, the TOST procedure also did not provide sufficient evidence for equivalence between AR (\textit{M} = 4.95, \textit{SD} = 0.99) and VR (\textit{M} = 4.74, \textit{SD} = 1.26), $\delta = 0.429$, \textit{p} = .09, corrected \textit{p} = .55. Thus, H5 was \textit{not supported}.
For \textbf{H6}, the Pearson correlation was small and negative, \textit{r} = -.12, and not statistically significant, \textit{p} = .73, corrected \textit{p} = 1.00. Therefore, H6 was \textit{not supported}. 
For \textbf{H7}, the Pearson correlation was close to zero, \textit{r} = .00, and not statistically significant, \textit{p} = .51, corrected \textit{p} = 1.00. Thus, H7 was \textit{not supported}.

Overall, \textit{only H3 was supported} after the Holm-Bonferroni correction. This difference between involvement in VR and involvement in AR is also clearly visible in \autoref{fig:result-ipq}.
%However, as stated above, no statistically significant differences were found for H1 or H2; no evidence for equivalence was found for H4 or H5; and no significant associations were observed for H6 or H7.

\subsection{Exploratory Comparisons Between Conditions}
\label{subsec:explo-tests}

\autoref{tab:stat_comparison} shows the results of the exploratory comparisons of the remaining subscales of all questionnaires. 
Data were parametric (normally distributed and with equal variances) in 10 cases, for which we used \textit{t}-tests. 
For three tests, we used Wilcoxon's signed-rank test instead.
As shown in \autoref{tab:stat_comparison}, we did not observe any significant differences among the four questionnaires between the AR and VR conditions.
For visual confirmation, we plotted medium values and standard deviations in \autoref{fig:result-fks}, \autoref{fig:result-ipq}, \autoref{fig:result-nmpsi}, and \autoref{fig:result-panas}.

\begin{table}[h!]
	\tiny
	\caption{Results for Exploratory Comparisons Between Conditions}
	\label{tab:stat_comparison}
	\resizebox{\linewidth}{!}{
		\begin{tabular}{lcclcc}
			\toprule
			\textbf{Condition}				& 	\textbf{AR} 		& \textbf{VR} 			& 				& &  \\\rowcolor{Grey}
			Variable      			&   \textit{M(SD)}  & \textit{M(SD)}   & \multicolumn{1}{c}{\textit{t}(df)/\textit{Z}$^\dagger$} &\textit{r}  &\textit{p}\\
			\midrule
			
			%PANAS: Positive Affect & 3.50(0.63) & 3.52(0.68) & \textit{t}(27) = -0.14 & .03 & .89 \\
			\rowcolor{Grey}
			PANAS: Negative Affect$^\dagger$ & 1.35(0.37) & 1.41(0.42) & \textit{Z} = -0.42 & -.08 & .68 \\
			NMSPI: Mutual Awareness & 4.12(0.36) & 3.97(0.30) & \textit{t}(27) = 1.60 & .29 & .12 \\
			\rowcolor{Grey}
			NMSPI: Attention & 3.40(0.82) & 3.26(0.71) & \textit{t}(27) = 0.98 & .19 & .33 \\
			NMSPI: Empathy & 3.90(1.11) & 3.99(1.04) & \textit{t}(27) = -0.74 & .14 & .47 \\
			\rowcolor{Grey}
			NMSPI: Mutual Understanding$^\dagger$ & 4.92(1.51) & 4.96(1.42) & \textit{Z} = -0.11 & -.02 & .91 \\
			NMSPI: Behavioral Interdependence & 5.27(1.01) & 5.18(1.15) & \textit{t}(27) = 0.59 & .11 & .56 \\
			\rowcolor{Grey}
			FKS: Flow Total & 4.95(0.99) & 4.74(1.26) & \textit{t}(27) = 1.28 & .24 & .21 \\
			FKS: Worry$^\dagger$ & 1.79(1.25) & 1.67(1.05) & \textit{Z} = 0.49 & .09 & .62 \\
			\rowcolor{Grey}
			FKS: Fluency & 4.87(1.18) & 4.70(1.39) & \textit{t}(27) = 0.90 & .17 & .37 \\
			FKS: Absorption & 5.06(1.06) & 4.80(1.28) & \textit{t}(27) = 1.38 & .26 & .18 \\
			\rowcolor{Grey}
			%CSI Score & 61.10(16.56) & 61.15(17.68) & \textit{t}(26) = -0.02 & .00 & .98 \\
			IPQ: General Presence & 4.54(1.93) & 4.36(1.77) & \textit{t}(27) = 0.39 & .07 & .70 \\
			%\rowcolor{Grey}
			IPQ: Spatial Presence & 4.66(0.97) & 4.75(1.02) & \textit{t}(27) = -0.45 & .09 & .66 \\
			%\textbf{IPQ: Involvement} & \textbf{3.55(0.85)} & \textbf{4.93(1.19)} & \textbf{\textit{t}(27) = -5.12} & \textbf{.70} & \textbf{\textless.001*} \\
			\rowcolor{Grey}
			IPQ: Realism & 3.38(0.84) & 2.95(1.10) & \textit{t}(27) = 1.97 & .36 & .06 \\
			
			\bottomrule
			$^\dagger$Wilcoxon test applied *\textit{p}\textless0.05
		\end{tabular}
	}
\end{table}

\subsection{Explorative Correlation Analysis}
\label{subsec:corr-analysis}
\begin{figure}[h!]
    \centering
    \includegraphics[width=.9\linewidth]{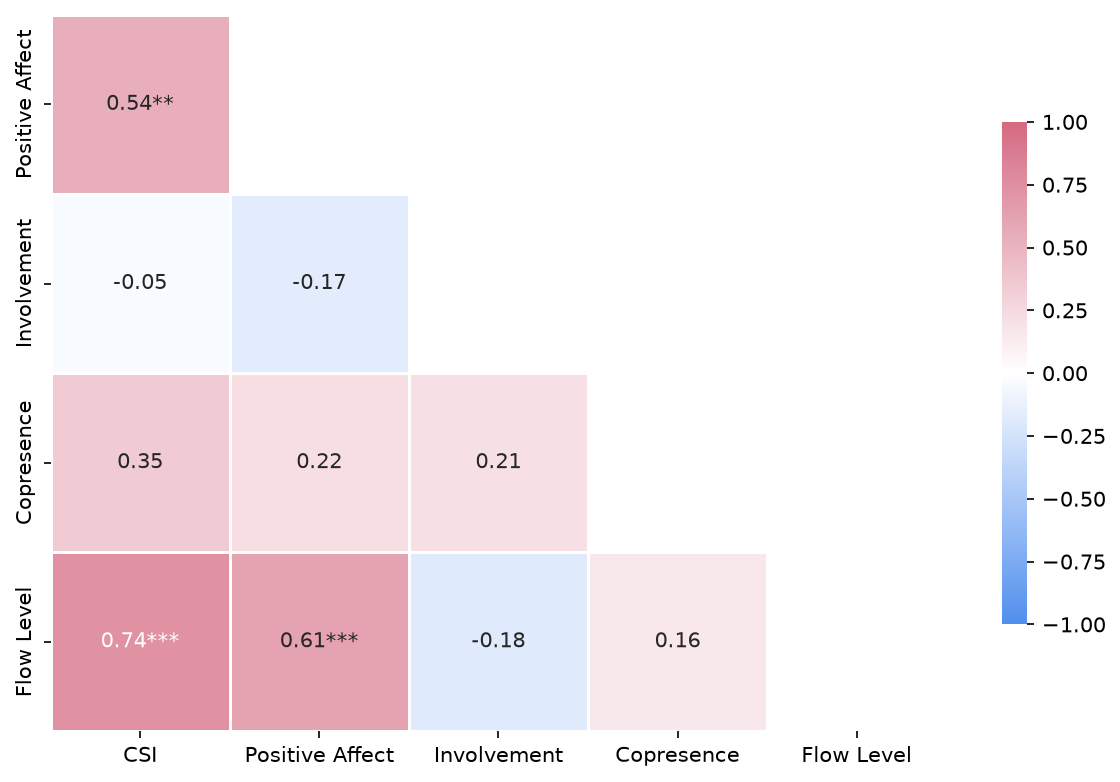}
    \caption{Pearson correlation coefficients in the AR condition.}
    \label{fig:correlation-ar}
\end{figure}

\begin{figure}[h!]
    \centering
    \includegraphics[width=.9\linewidth]{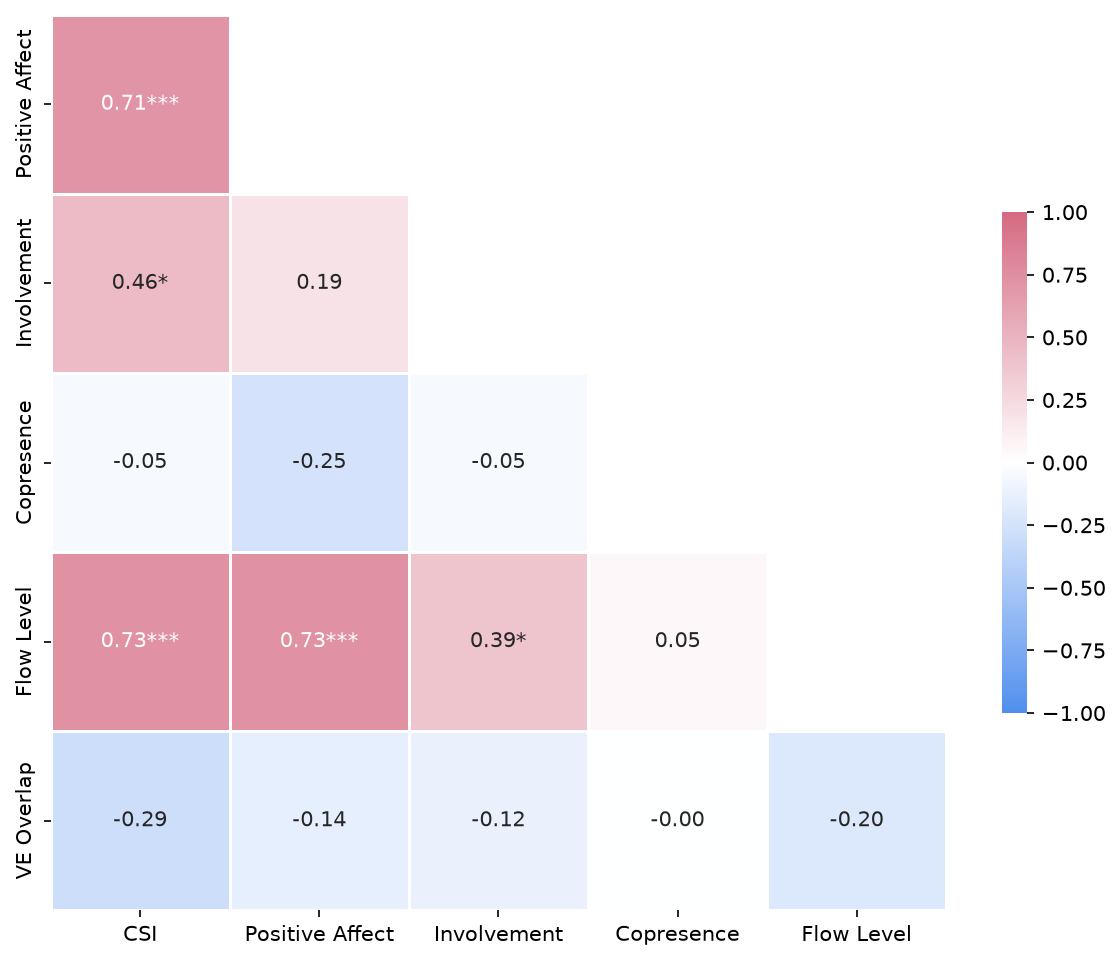}
    \caption{Pearson correlation coefficients in the VR condition.}
    \label{fig:correlation-vr}
\end{figure}

The results are plotted in two separate correlation matrices in \autoref{fig:correlation-ar} and \autoref{fig:correlation-vr}.
For both AR and VR, we observed \textit{strong and significant correlations among the positive affect, flow, and creativity support scales} (5 out of 6 correlations with \textit{p} $< .001$ and \textit{p} $< .01$ for CSI vs. positive affect in AR).
For VE overlap and copresence, however, we did not see any significant correlations, neither in AR nor in VR.
Exclusively in VR, we observed \textit{moderate significant correlations (p $< .05$)  between involvement and flow, and between involvement and creativity support} (see \autoref{fig:correlation-vr}).

\subsection{Interview Analysis}
\label{subsec:quali-analysis}

\begin{figure}[hb!]
    \centering
    \includegraphics[width=.95\linewidth]{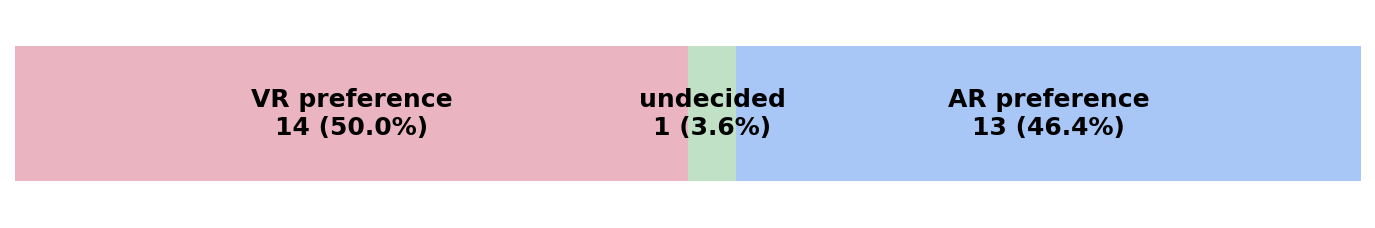}
    \caption{Participants' post-test preferences for jamming in VR or AR.}
    \label{fig:preference}
\end{figure}

% --- AR / VR preference reasons ---
\begin{figure*}[htbp]
    \centering

    \begin{minipage}[t]{0.46\textwidth}
        \centering
        \includegraphics[width=.87\linewidth]{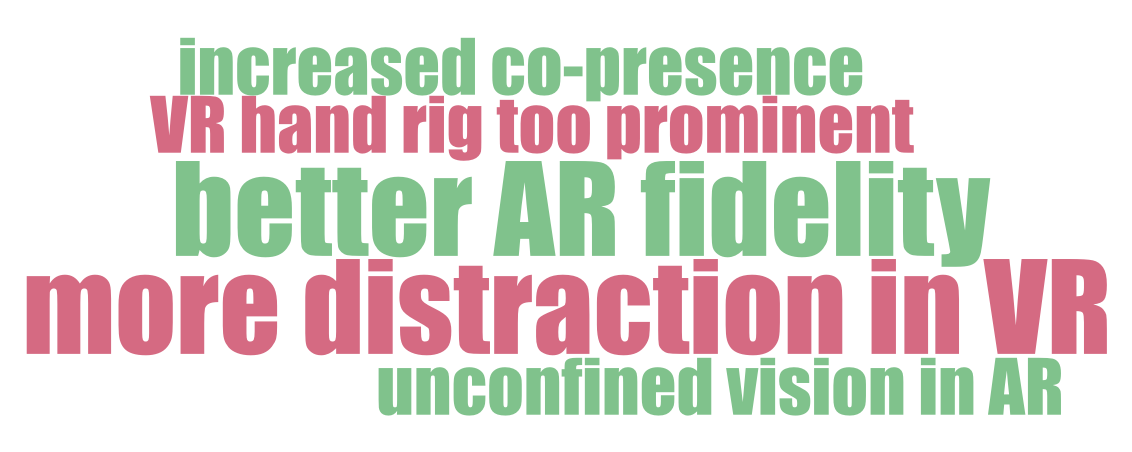}
        \caption{A code cloud showing the most frequently coded reasons for AR preference stated by participants.}
        \label{fig:ar-preference-reasons}
    \end{minipage}
    \hspace{0.03\textwidth}
    \begin{minipage}[t]{0.46\textwidth}
        \centering
        \includegraphics[width=.87\linewidth]{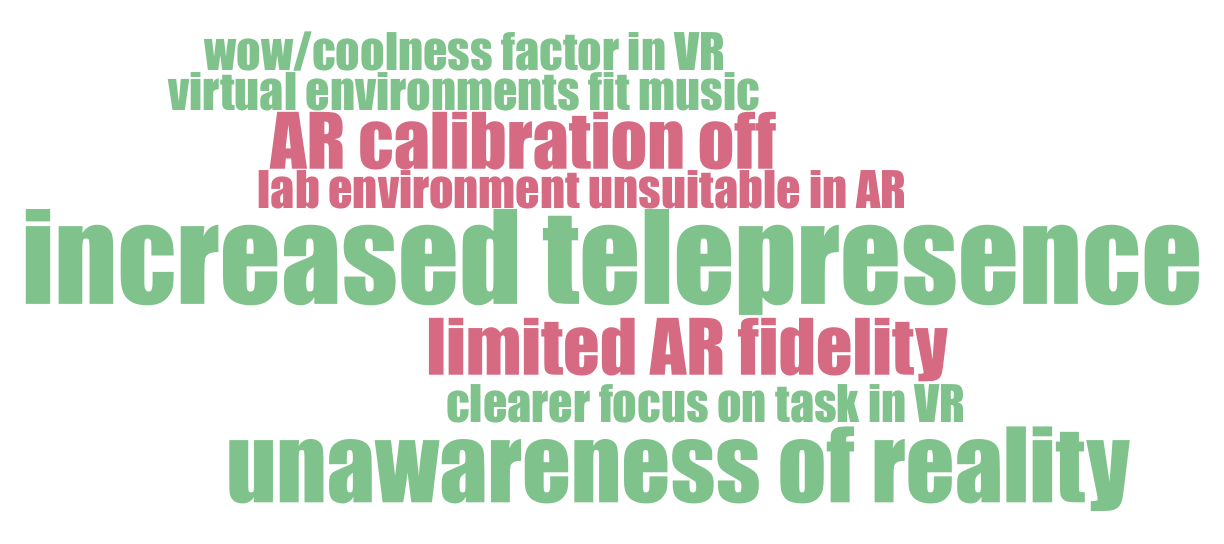}
        \caption{A code cloud showing the most frequently coded reasons for VR preference stated by participants.}
        \label{fig:vr-preference-reasons}
    \end{minipage}

\end{figure*}

% --- Creative environment preferences ---
\begin{figure*}[htbp]
    \centering

    \begin{minipage}[t]{0.46\textwidth}
        \centering
        \includegraphics[width=.87\linewidth]{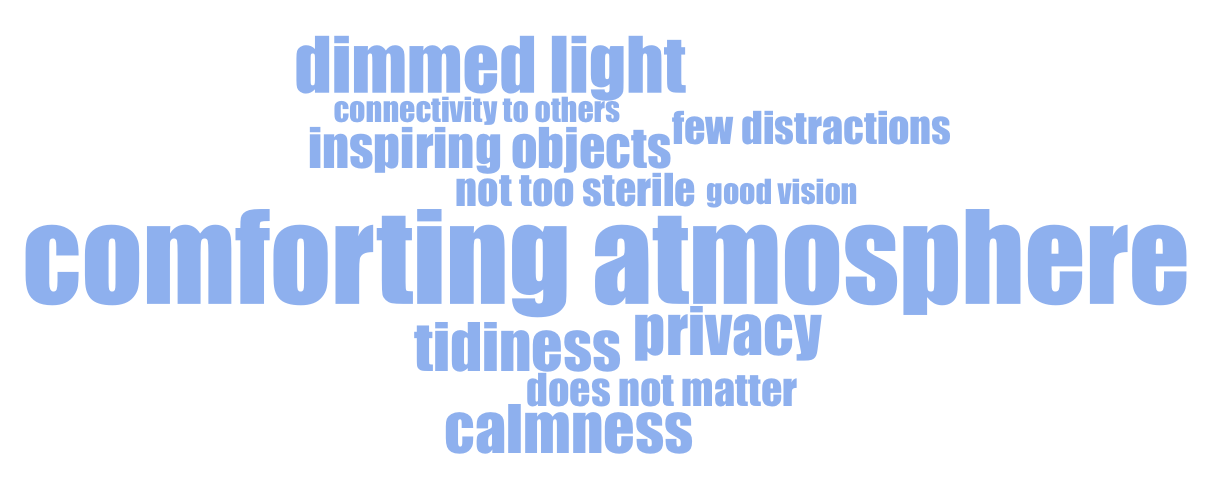}
        \caption{A code cloud showing the most frequently coded requirements for creative environments.}
        \label{fig:creative-preferences}
    \end{minipage}
    \hspace{0.03\textwidth}
    \begin{minipage}[t]{0.46\textwidth}
        \centering
        \includegraphics[width=.87\linewidth]{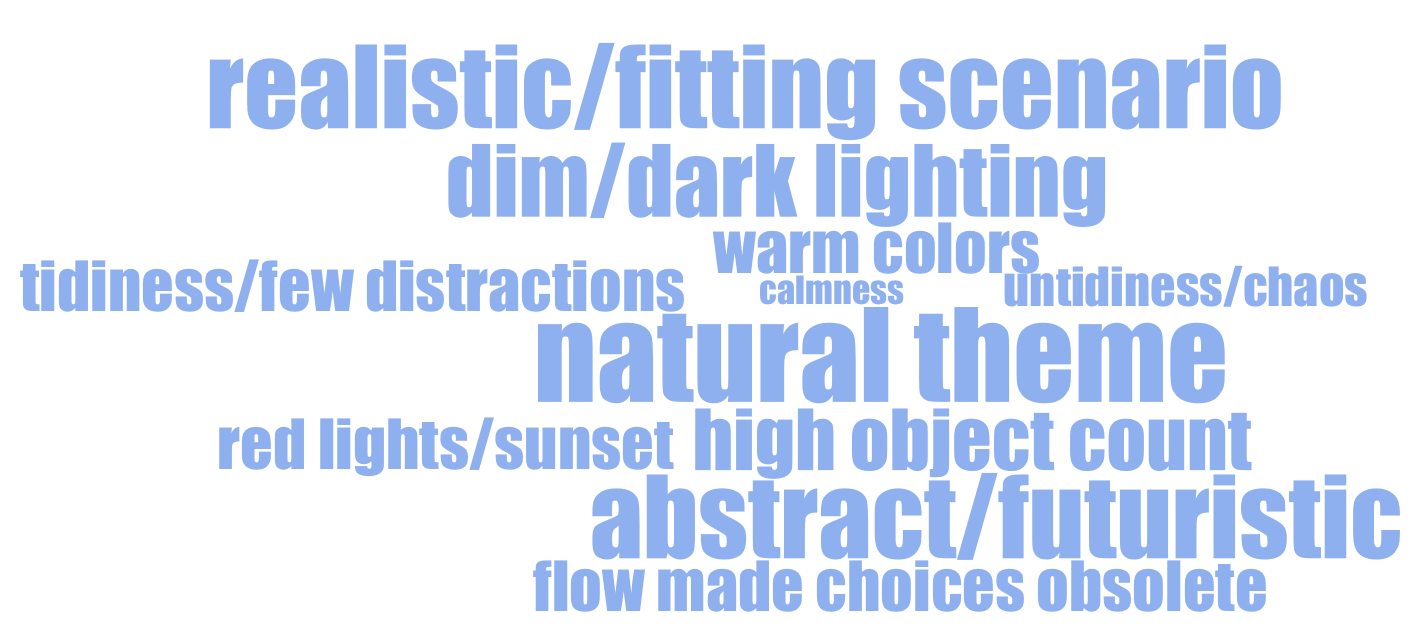}
        \caption{A code cloud showing the most uttered priorities for selection and customization of virtual environments in VR.}
        \label{fig:env-choices}
    \end{minipage}

\end{figure*}

\subsubsection{Preference for AR or VR (RQ3)}

After participants had experienced one jam session in AR and one in VR, we asked which condition they preferred and why.
Preferences were nearly evenly distributed (see \autoref{fig:preference}), with 14 participants (50\%) preferring VR, 13 (46.4\%) preferring AR, and one participant being undecided.
Participants justified their choices both by referring to the strengths of their preferred condition and by criticizing the alternative condition.
In \autoref{fig:ar-preference-reasons} and \autoref{fig:vr-preference-reasons}, red codes indicate critical aspects of the non-preferred condition, while green codes indicate positive aspects of the preferred condition.
Overall, reasons for preference varied widely, with the most frequent codes occurring only three to five times.

\subsubsection{Reasons for AR preference (RQ3)}

Participants who preferred AR most often cited higher perceived environmental fidelity and fewer distractions compared to VR, each mentioned three times (see \autoref{fig:ar-preference-reasons}).
Unconfined vision, increased co-presence, and the overly prominent VR hand rig were each mentioned twice.
Overall, AR was associated with \textit{visual fidelity, openness, and a closer alignment with expectations}, whereas \textit{VR was criticized for distraction, reduced orientation, and confined vision}.

\subsubsection{Reasons for VR preference (RQ3)}

Participants who preferred VR most often emphasized increased telepresence, mentioned five times, followed by reduced awareness of the physical laboratory environment, mentioned four times (see \autoref{fig:vr-preference-reasons}).
AR calibration issues and limited AR fidelity, particularly the screendoor effect, were each mentioned three times.
Further recurring reasons included that the virtual environments fit the music-making task, that the laboratory environment was unsuitable in AR, that VR enabled clearer focus on the task, and that VR had a stronger wow or coolness factor, each mentioned twice.
Overall, \textit{VR preference was mainly explained by immersion, telepresence, and the ability to block out an unsuitable physical environment}, while \textit{AR was criticized for calibration problems, limited fidelity, and insufficient virtual content}.

\subsubsection{Requirements for Creative Environments (RQ4)}

%For RQ4, we first asked for general requirements for creative spaces independently of XR; participants most frequently emphasized the need for a cozy or comforting atmosphere, mentioned by 11 participants (see \autoref{fig:creative-preferences}).
%More specific descriptions of this atmosphere included dimmed lighting (mentioned six times) and calmness, privacy, and tidiness (each mentioned five times).
%Inspiring objects, such as images, scenery, or specific furniture, were mentioned four times.
%Less frequently, participants stated that the surroundings did not matter while making music, that the space should not be too distracting, or that it should not feel too sterile, each mentioned three times.
%Good visibility and connectivity with others were mentioned twice.
%Overall, participants emphasized that creative spaces should provide a \textit{cozy, calm, private, and inspiring atmosphere that supports focus} while \textit{avoiding excessive distraction or sterility}.
For RQ4, we first asked about general requirements for creative spaces independently of XR. Participants most frequently emphasized a cozy or comforting atmosphere (11 participants; see \autoref{fig:creative-preferences}), including dimmed lighting (six), calmness, privacy, and tidiness (five each). Inspiring objects, such as images, scenery, or specific furniture, were mentioned four times. Participants less frequently stated that surroundings did not matter, should not be too distracting, or should not feel too sterile (three each), while good visibility and connectivity with others were mentioned twice. Overall, participants emphasized a \textit{cozy, calm, private, and inspiring atmosphere that supports focus} while \textit{avoiding excessive distraction or sterility}.

%------------------ Q3 (VE choices) ------------------
\subsubsection{VE Customization Choices (RQ4)}

Creative needs were only partially reflected in participants' VE choices and customizations (see \autoref{fig:env-choices}).
Preferences clustered around three environment types: natural settings such as the `meadow' environment, abstract or futuristic settings such as `space', and realistic music-related settings such as `theater' or `studio'.
Customization priorities were mainly concerned with dim or dark lighting, warm colors, tidiness, visual complexity, and, for some participants, disorder or chaos.
Several participants also stated that the environment became less relevant once they entered a flow state.
Overall, VE preferences varied widely, suggesting that customizable environments should support diverse needs rather than assume a single ideal creative space.

\section{Discussion}

\paragraph{On the interconnectedness of creativity support, positive affect, and flow in XR}

%We found strong correlations between creativity support, positive affect, and flow in both AR and VR conditions.
%This suggests that these constructs may not be fully independent in the context of XR-supported music-making.
%Instead, they may reflect a more general experiential factor that captures the overall quality of the XR jamming experience.
%Future work could investigate this possibility through a dedicated factor analysis.
We found strong correlations between creativity support, positive affect, and flow in both AR and VR, suggesting that these constructs may reflect a more general experiential factor capturing the overall quality of the XR jamming experience. Future work could investigate this possibility through factor analysis.

\paragraph{VR-based involvement is not universally appreciated}

Participants reported a significantly stronger sense of involvement in VR, suggesting that the customizable VEs successfully increased immersion in the respective scenes.
Here, involvement refers to engagement with the immersive environment rather than with the musical task itself.
However, this increased involvement did not yield significant benefits for creativity support, affect, flow, or co-presence.
Based on the qualitative data, we interpret this as a result of substantial individual differences in participants' preferences for AR and VR.
While some participants appeared to feel more creatively supported in AR, others seemed to benefit more from VR, potentially causing these effects to cancel out at the group level.
This interpretation is supported by the nearly even preference distribution: 14 participants preferred VR, 13 preferred AR, and only one was undecided (see \autoref{fig:preference}).
This polarization may also explain why we did not observe correlations with proportional time spent in the same virtual environments (i.e., VE overlap).
If only part of the sample experienced VR as creatively beneficial, sharing the same or different VEs may not have produced a consistent effect across participants.

\paragraph{Involvement has correlations exclusive to VR}

% Involvement showed medium correlations with both flow and creativity support, but only in the VR condition.
% This finding further supports the assumption that participants with stronger immersive tendencies or a preference for VR experienced more creativity support and better flow in VR. 
% In contrast, participants who preferred AR may not have benefited from increased involvement in the same way, but may rather have felt increasingly distracted from making music.
% However, we did not measure potentially contributing factors like immersive tendencies (e.g., with the ITQ \cite{witmer1998measuring}), prior XR experience, or motion-sickness sensitivity in our study. 
% Accordingly, future studies should aim to identify such contributing factors to AR or VR preference. 
% Similarly, we did not examine how diads that know each other versus diads that just met perform in either condition, which could be a topic for future studies. 
% Furthermore, we note that significant correlations in only one condition do not necessarily imply significant differences between conditions for correlating measures.
Involvement showed medium correlations with both flow and creativity support, but only in VR, suggesting that participants with stronger immersive tendencies or a preference for VR may have experienced greater creativity support and flow, whereas those preferring AR may have experienced increased involvement as distracting from music making. 
However, we did not measure potentially contributing factors such as immersive tendencies (e.g., with the ITQ \cite{witmer1998measuring}), prior XR experience, or motion-sickness sensitivity, which future studies should examine as potential contributors to AR/VR preference. 
Future work should also investigate whether familiarity between dyad members affects performance in either condition. Importantly, correlations observed in only one condition do not necessarily imply significant between-condition differences in the correlating measures.

\paragraph{Different aspects are perceived as distracting}

Participants differed substantially in which environmental properties supported or distracted from the musical task. While some found VR distracting, others reported that its increased immersion helped them focus. 
Similarly, preferences varied across natural, abstract, futuristic, and realistic environments, although dimmer lighting and control over visual distraction emerged as recurring customization goals. 
These findings support customizable rather than one-size-fits-all VEs, while such customization may also introduce additional variability into experimental comparisons.
%Participants' explanations of their AR and VR preferences suggest that focusing on the musical task was a central goal, but that different aspects were experienced as distracting.
%While some participants felt more distracted in VR, others reported that VR's increased presence and immersion helped them focus on making music.
%Likewise, environment selection and customization showed no clearly dominant preference, as participants variously favored natural, abstract, futuristic, and realistic music-related environments.
%The most consistent customization goals concerned dimmer lighting and the regulation of visual distraction.
%Thus, VR customization supported focus for some participants, but the environmental properties that enabled this focus varied substantially among individuals.
%As such, we think giving participants the freedom to select and customize VEs was beneficial, even though it may have made individual VR experiences more unique and introduced potential confounding variables, making significant effects less likely.

\paragraph{Additional notes}

% Four participants also stated that flow made their VE customization choices less relevant, as focusing on the musical task made them forget their surroundings.
% While this aligns with previous observations and theoretical assumptions regarding flow \cite{anonymous2026system}, the finding remains difficult to interpret because only a minority mentioned it and we did not ask about it directly in the interview.
% As an additional observation, positive affect increased, and negative affect decreased significantly after jamming in both conditions (see \autoref{fig:result-panas}), which is in line with previous work \cite{anonymous2023system}.
% However, due to space limitations, we do not report these results in detail here.
Four participants reported that flow made VE customization less relevant as focusing on the musical task reduced awareness of their surroundings. While consistent with previous observations and flow theory \cite{schlagowskiMusicalCreativityFlow2026}, this finding is difficult to interpret because only a minority mentioned it and it was not directly elicited. Additionally, positive affect significantly increased and negative affect decreased after jamming in both conditions (see \autoref{fig:result-panas}), consistent with previous work \cite{schlagowskiWishYouWere2023}, although space limitations preclude detailed reporting.

\paragraph{Limitations}

One limitation is the potential dependency between dyads, as participants jammed in pairs. However, participants made individual VE customization choices and completed questionnaires separately in different rooms; we therefore treated them as independent samples while using dependent-measures tests for within-participant comparisons. 
The 20-minute jam sessions were also relatively short, with some participants reporting that sessions ended while they were in a flow state, although this indicates that they were sufficient to elicit measurable flow. 
Together, the short sessions, unfamiliar laboratory environment, limited sample size, and potential dyadic dependencies may have reduced our ability to detect effects beyond H3.
Technical limitations included occasional inaccuracies in VR hand tracking. 
The hand rig introduced to improve fingertip visibility distracted some participants, and both it and the 3D UI were exclusive to VR, potentially affecting measures such as CSI and counteracting benefits of customizable environments.
Fidelity was criticized in both conditions, particularly VE realism in VR and video pass-through quality in AR. 
Point-cloud criticism mainly concerned calibration, especially the more visible misalignment with physical objects in AR, whereas point-cloud resolution was rarely criticized, suggesting that the Gaussian Splatting shader performed adequately. 
HMDs also obscured participants’ faces, limiting gaze-based communication, which some musicians considered important for jamming.
Finally, the German NMSPI version was not psychometrically validated, although carefully translated and adapted from previous work. The male-skewed sample (22 of 28 participants) further limits generalizability.

\section{Conclusion}

In this work, we investigated how immersive, customizable Virtual Environments (VEs) affect networked music performance compared with participants’ physical surroundings in AR. 
VR increased involvement, suggesting greater immersion, but provided no clear advantages in creativity support, affect, co-presence, or flow (RQ1). 
Psychometric outcomes showed no consistent relationships with time spent in similar or different VEs, suggesting no systematic effect of environment overlap on individual jamming experiences (RQ2). 
Preferences were almost evenly split: some participants valued VR for immersion, focus, and engaging environments, whereas others preferred AR for lower perceived distraction, greater familiarity, and closer alignment with real-world interaction (RQ3). 
Our customizable VR apparatus addressed some participants’ needs through dim lighting, customizable scenery, and control over visual stimulation, but not universally due to individual preferences and technical limitations (RQ4).
Overall, our findings suggest that the surrounding environment affects collaborative musical experiences differently depending on individual preferences for immersion and situatedness. Based on the observed correlations, VR may support creativity and flow for musicians who benefit from immersive environments, whereas AR may better suit those preferring less mediated, reality-anchored experiences. Given the substantial variation in environmental needs indicated by both quantitative and qualitative findings, NMP applications for the musical metaverse should therefore support a broad range of the reality-virtuality continuum to cover individual needs and preferences.

\bibliographystyle{IEEEtran}
\bibliography{IS2}

\end{document}